\documentclass[]{pasj01}
\twocolumn

\begin{document} 
\Received{}%{yyyy/mm/dd}
\Accepted{}%{yyyy/mm/dd}
%\Published{yyyy/mm/dd}

\title{The UV spectral slope beta and stellar population of most active star-forming galaxies at z$\sim$4}

%%% begin:list of authors
% Do NOT capitalize all letters in "textsc".
\author{Satoshi \textsc{Yamanaka}\altaffilmark{1}\altaffilmark{2}\altaffilmark{3}}
\altaffiltext{1}{Department of Environmental Science and Technology, Faculty of Design Technology, Osaka Sangyo University, 3-1-1, Nakagaito, Daito, Osaka 574-8530, Japan}
\altaffiltext{2}{National Astronomical Observatory of Japan, 2-21-1 Osawa, Mitaka, Tokyo 181-8588}
\altaffiltext{3}{Astronomical Institute, Tohoku University, Aramaki, Aoba-ku, Sendai, Miyagi 980-8578, Japan}
\email{s.yamanaka@est.osaka-sandai.ac.jp}

\author{Toru \textsc{Yamada},\altaffilmark{4}}
\altaffiltext{4}{Institute of Space and Astronautical Science, Japan Aerospace Exploration Agency, 3-1-1, Yoshinodai, Chuo-ku, Sagamihara, Kanagawa, 252-5210, Japan}
\email{yamada@ir.isas.jaxa.jp}
%%% end:list of authors

%% `\KeyWords{}' always has to be placed before `\maketitle'.
 %Do NOT move this preamble from here!

\maketitle

\begin{abstract}
  
 We investigate the stellar population of star-forming galaxies at \textit{z} $\sim$ 4 by focusing on their slope of rest-frame ultraviolet (UV) continuum, $\beta$ where $f_{\lambda} \propto \lambda^{\beta}$. We investigate the sample of bright Lyman Break Galaxies (LBGs) with \textit{i'} $\leq$ 26.0 in the Subaru/XMM-Newton Deep Survey field by using the SED fitting analysis.
 We find that the apparently redder ($\beta_{\mathrm{obs}} > -1.73$) LBGs tend to be dusty ($\mathrm{Av} > 1.0$), have the young stellar population ($\beta_{\mathrm{int}} < -2.42$), and the intrinsically active star-forming galaxies (SFR $\gtrsim$ a few $\times\ 10^{2}\,\mathrm{M_{\solar}}\>\mathrm{yr^{-1}}$). It means that a significant fraction of the UV--selected LBGs at \textit{z} $\sim$ 4 is on-going active and dust obscured star-forming galaxies.
 We compare the IR to UV luminosity ratio assuming the dust attenuation laws with the sub-millimeter observations from previous works. The result suggests that the Calzetti-like dust attenuation law is preferable for the active and dusty star-forming LBGs at $z = 4$.
 We also find that an extrapolation of the $\beta_{\mathrm{int}}$--$M_{\mathrm{UV,int}}$ relation toward the fainter magnitude range below our sample magnitude limit intersects the $\beta_{\mathrm{obs}}$--$M_{\mathrm{UV,obs}}$ relation previously obtained in the deeper narrow-area observations at $M_{\mathrm{UV}} = -18.9$ and $\beta = -1.94$, which coincides with the break point of $\beta_{\mathrm{obs}}$--$M_{\mathrm{UV,obs}}$ relation observed so far. The coincidence suggest that we see the almost dust-free population at $M_{\mathrm{UV,obs}} \gtrsim -18.9$.
 
\end{abstract}

%%%%%%%% 1. Introduction %%%%%%%%
\section{Introduction} \label{S1Intr}

 The ultraviolet (UV) continuum spectrum of star-forming galaxies is characterized by the spectral index $\beta$ in the form of $f_{\lambda} \propto \lambda^{\beta}$ \citep{Calz94}. The $\beta$ values are related with physical quantities such as age, metallicity, and dust extinction of the galaxies. In the case of less dust attenuation, younger age, and lower metallicity, the galaxy has the larger negative $\beta$ value (e.g., \cite{Bouw10}; \cite{Stan16}). Since it is relatively simple to measure $\beta$ values even if objects are at high redshift, the $\beta$ index is a useful tool to probe their physical quantities.

 The typical value of $\beta$ found in the previous works is $\sim -1.7$ at $z \sim 4$ for $\sim L_{*}$ galaxies with $M_{\mathrm{UV}} \sim -21.0$ (\cite{Bouw12}; \cite{Fink12}), and it is bluer at higher redshift up to $z \sim 7$ ($\beta$--\textit{z} relation: e.g., \cite{Wil11}). At the fainter magnitude (e.g. $M_{\mathrm{UV}} \sim -19.0$), the observed $\beta$ value has still uncertainties and the relation between $\beta$ and UV absolute magnitude ($\beta$--$M_{\mathrm{UV}}$ relation) has been a subject of debate for the several years (e.g., \cite{Bouw12}; \cite{Bouw14}; \cite{Dunl12}; \cite{Dunl13}; \cite{Fink12}; \cite{Rog14}). \authorcite{Bouw12} and \authorcite{Rog14} report that bright galaxies have redder $\beta$ values and faint galaxies have bluer $\beta$ values, while \authorcite{Dunl13} and \authorcite{Fink12} report that $\beta$ values are constant over the observed magnitude range. This inconsistency in the $\beta$--$M_{\mathrm{UV}}$ relation can be caused by both large photometric errors for faint galaxies and selection bias. In order to reveal the true $\beta$--$M_{\mathrm{UV}}$ relation, a further large sample of objects with small photometric uncertainties is needed and/or it is necessary to assess the incompleteness of the observed $\beta$ distribution. Recently, \citet{Dunc15} and \citet{Bouw14} show that the $\beta$ value decreases with the $M_{\mathrm{UV}}$ value. \authorcite{Dunc15} finds the trend by combining the results of the previous literature (\cite{Bouw14};\cite{Dunc14}; \cite{Dunl12}; \cite{Dunl13}; \cite{Fink12}; \cite{Rog14}; \cite{Wil11}) and \authorcite{Bouw14} discusses the trend by assessing the observational bias (incompleteness) of the observed $\beta$ distribution in the faint magnitude range.

 The $\beta$--$M_{\mathrm{UV}}$ relation is understood as another aspect of the mass-metallicity relation seen in star-forming galaxies as the $\beta$ value depends on the dust extinction (\cite{Bouw09}, \yearcite{Bouw12}, \yearcite{Bouw14}). Moreover, it is suggested that there is a ``knee'' in the $\beta$--$M_{\mathrm{UV}}$ relation at $M_{\mathrm{UV}} \sim -19.0$, and the dependence of $\beta$ on $M_{\mathrm{UV}}$ becomes weaker at $M_{\mathrm{UV}} \gtrsim -19.0$ than at $M_{\mathrm{UV}} < -19.0$ \citep{Bouw14}. The change of the dependence is interpreted as the change of the dependence of the dust extinction on the UV luminosity or the stellar mass (e.g., \cite{Pann09}; \cite{Redd10}). The semi-analytic model predicts the sudden change of the dust-to-gas mass ratio at the critical metallicity and the dust mass rapidly increase at the metallicity level larger than the critical metallicity (e.g., \cite{Hira11}). The change of the dependence of $\beta$ on $M_{\mathrm{UV}}$ perhaps indicates the existence of the critical metallicity. The redshift dependence of $\beta$ is interpreted as the dust attenuation history or the dust production history in star-forming galaxies. Interestingly, the dust attenuation history at $z \gtrsim 3$ estimated from the redshift dependence of $\beta$ is smoothly connected with the dust attenuation history at $z \lesssim 3$ estimated from the direct measurements of both IR and UV luminosity \citep{Burg13}. The dust attenuation history is also used for revealing the history of the true (dust-corrected) cosmic Star Formation Rate (SFR) density in the high-z universe because it is still difficult to obtain the IR luminosity for high-z star-forming galaxies (e.g., \cite{Bouw09}, \yearcite{Bouw12}; \cite{Mada14}). Currently, the $\beta$--\textit{z} relation has been used for considering the source of cosmic reionization, assuming that the $\beta$ value represents the production rate of hydrogen ionizing photons since the production rate is susceptible to the stellar population (e.g., \cite{Dunc15}; \cite{Bouw15}, \yearcite{Bouw16a}).
 
 We note that the samples in most of the literature are overlapped and not independent since the set of GOODS-South/HST or HUDF/HST data is mostly used so far in their works. Due to the small observed area, the number of bright objects at $z = 4$ in the field is limited (except for \cite{Rog14}) and these previous studies focus on relatively faint galaxies at high redshift. The $\beta$ distribution of luminous objects is, however, also important since such population provides important clues to understand early star-formation history in the universe (e.g., \cite{Cucc12}; \cite{Hatf18}). In fact, by using stacked images, \citet{LeeKS11} investigate the $\beta$ value of ultra-luminous star-forming galaxies ($19.46 \leq I < 24.96$) at \textit{z} $\sim$ 4 and they find that the $\beta$ value of the stacked star-forming galaxies tends to be redder toward the brighter magnitude range. When taking all the previous works into consideration, the investigation for individual and ''normal'' luminous galaxies ($L \sim L_{*}$) is very comprehensive.

 Recent Atacama Large Millimeter/submillimeter Array (ALMA) observations have revealed the dust properties of high redshift star-forming galaxies through the relation between the ratio of IR to UV (so called IRX) and the UV slope $\beta$ (IRX--$\beta$ relation; e.g., \cite{Cpk15}; \cite{Bouw16}; \cite{McLu18}). In order to interpret these results, it is very important to investigate the detailed relation between the UV slope $\beta$ and the stellar population which is hidden by dust extinction. On the other hand, there is difficulty in studying the intrinsic values of the $\beta$, $M_{\mathrm{UV}}$, and stellar population for the luminous and massive galaxies, since the effect of dust extinction on their color degenerates with the age and metallicity of their stellar population: the more massive systems are on average older and metal rich. It is essential to resolve these degeneracy and understand the intrinsic properties of star formation and effects of dust extinction.
 
 In this paper, we present the results of the UV slope $\beta$ and stellar population for the relatively bright galaxies with $M_{\mathrm{UV}} \lesssim -20$ at $z \sim 4$ in relatively wide-area and deep Subaru/XMM-Newton Deep Survey (SXDS) field. Wide area coverage is essential to sample rare luminous galaxies. In section 2, we explain the data and sample selection. In section 3, we describe our method to evaluate the $\beta$ value. In section 4, we show the result of the observed $\beta$--$M_{UV}$ relation, and we assess the incompleteness of our sample selection. In section 5, we discuss the intrinsic $\beta$--$M_{UV}$ relation, most active star-forming galaxies, and dust attenuation law for $z \sim 4$ galaxies. In section 6, we give our conclusions to summarize this work. In regard to the cosmological parameters, we assume $\Omega_{m,0} = 0.3$, $\Omega_{\Lambda,0} = 0.7$, $H_{\mathrm{0}} = 70\>\mathrm{km\>s^{-1}\>Mpc^{-1}}$. Finally throughout this work we apply the AB magnitude system (\cite{OkeGunn}; \cite{Fkgt96}).

%%%%%%%% 2. Data & Sample %%%%%%%%
\section{ Data and sample selection } \label{S2Dtss}

%%%%%%%% 2.1 Data %%%%%%%%
\subsection{ Data } \label{S2s1da}

\begin{figure}
  \begin{center}
    \includegraphics[width=80mm]{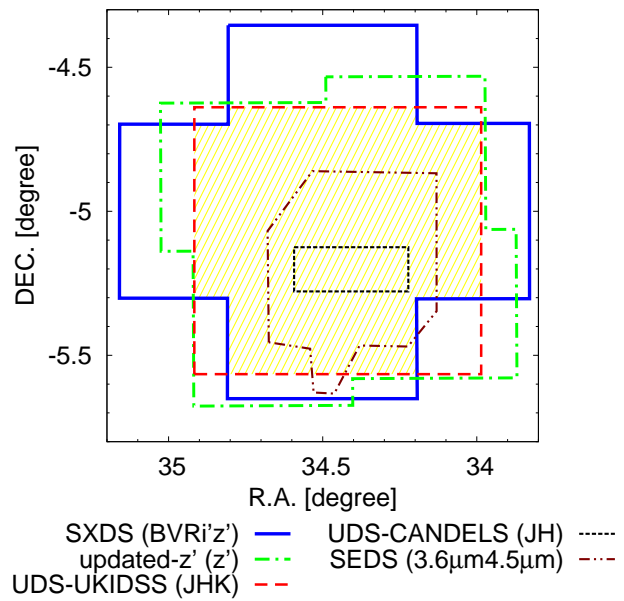}
  \end{center}
  \caption{ Field map of imaging data used in our analysis. The blue solid line show the sky coverage of the SXDS field; the green dot-dash line represent the sky coverage observed with Subaru/updated-\textit{z'}; The red dash line represents the sky coverage of the UDS-UKIDSS field; the black dotted line represents the sky coverage of the UDS-CANDELS field; and the brown two-dot-dash line represents the sky coverage of the SEDS field. Our catalog consists of the objects within the area covered by all of the SXDS, updated-\textit{z'}, and UDS field since we use the \textit{i'}, \textit{z'}, updated-\textit{z'}, and \textit{J} band photometry for estimating the UV slope $\beta$ value. This area is filled by the yellow slanting lines. } \label{fig1}
\end{figure}

 In our analysis, we select the Lyman Break Galaxies (LBGs) at $z \sim 4$ in the SXDS field which is partially covered by other surveys (i.e., UDS-UKIDSS/UKIRT, UDS-CANDELS/HST, and SEDS/Spitzer). Figure \ref{fig1} shows the field map of each survey and indicates that UDS-CANDELS and SEDS survey do not cover the entire field of SXDS or UDS-UKIDSS.

 Our catalog includes \textit{B}, \textit{V}, \textit{R}, \textit{i'}, \textit{z'}, and updated-\textit{z'} from Subaru/Suprime-Cam, \textit{J}, \textit{H}, and \textit{K} from UKIRT/WFCAM, F125W and F160W from HST/WFC3, and 3.6$\>\micron$ and 4.5$\>\micron$ from Spitzer/IRAC. The \textit{B}, \textit{V}, \textit{R}, \textit{i'}, \textit{z'} band images are taken from the archived SXDS data \citep{Furu08}. The CCD of Subaru/Suprime-Cam is replaced with the new CCD of Hamamatsu Photonics in 2008, and the total response function improve especially at the longer wavelength. The Subaru/Suprime-Cam \textit{z'} band images are taken after the replacement again, and we call the images the updated-\textit{z'} band. The limiting magnitude of the updated-\textit{z'} images is $\sim 0.5\>\mathrm{mag}$ deeper than the archived SXDS data \citep{Furu16}. The \textit{J}, \textit{H}, and \textit{K} band images are taken from the UKIRT Deep Sky Survey (UKIDSS: \cite{Law07}) DR10, and the F125W and F160W band images are taken from the Cosmic Assembly Near-Infrared Deep Extragalactic Legacy Survey (CANDELS: \cite{Grog11}; \cite{Koek11}). For the data from \textit{B} to \textit{K} band, we use 2''-diameter aperture magnitude measured by using the SExtractor\footnote{$\langle$http://www.astromatic.net/software/sextractor$\rangle$} ver.2.5.0 \citep{BeAr96}. The smaller PSF images are convolved with Gaussian kernels to be matched in FWHM of the stars (1''.0) with the original updated-\textit{z'} band image. On the other hand the 3.6$\>\micron$ and 4.5$\>\micron$ photometry are taken from the Spitzer Extended Deep Survey (SEDS) catalog \citep{Ash13} and we apply aperture correction. Finally, we pick up the objects which are in the overlapped region covered by both the SXDS and UDS-UKIDSS fields (see figure \ref{fig1}: the overlapped region is filled by yellow slanting lines) because we need both optical and NIR photometry for estimating the UV slope $\beta$ value of \textit{z} $\sim$ 4 galaxies. The information of the imaging data is summarized in table \ref{tab1}.

%--------- Figure1 ---------%
%--------- Figure1 ---------%

%---------- Table1 ----------%
\begin{table*}
  \tbl{Summary of survey data.}{%
    \begin{tabular}{lllccc}
      \hline
      Field Name & Instrument & Band & Limiting Mag. & PSF FWHM & Reference\footnotemark[$*$] \\
      (Survey Name) & &  & ($5\,\sigma$, 2''$\phi$ , AB) & (arcsec) & \\
      \hline
      SXDS & Subaru/Suprime-Cam & \textit{B} & 27.5 & 0.8 & (1) \\
      SXDS & Subaru/Suprime-Cam & \textit{V} & 27.1 & 0.8 & (1) \\
      SXDS & Subaru/Suprime-Cam & \textit{R} & 27.0 & 0.8 & (1) \\
      SXDS & Subaru/Suprime-Cam & \textit{i'} & 26.9 & 0.8 & (1)\\
      SXDS & Subaru/Suprime-Cam & \textit{z'} & 25.8 & 0.7 & (1) \\
      & Subaru/Suprime-Cam & updated-\textit{z'} & 26.5 & 1.0 & (2) \\
      UDS-UKIDSS & UKIRT/WFCAM & \textit{J} & 25.5 & 0.8 & (3) \\
      UDS-UKIDSS & UKIRT/WFCAM & \textit{H} & 24.9 & 0.8 & (3) \\
      UDS-UKIDSS & UKIRT/WFCAM & \textit{K} & 25.2 & 0.8 & (3) \\
      UDS-CANDELS & HST/WFC3 & F125W & 25.6 & 0.12 & (4), (5) \\
      UDS-CANDELS  & HST/WFC3 & F160W & 25.6 & 0.18 & (4), (5) \\
      SEDS & Spitzer/IRAC & 3.6$\>\micron$ & 24.75\footnotemark[$\dag$] & 1.8 & (6) \\
      SEDS & Spitzer/IRAC & 4.5$\>\micron$ & 24.8\footnotemark[$\dag$] & 1.8 & (6) \\
      \hline
  \end{tabular}}\label{tab1}
  \begin{tabnote}
    \footnotemark[$*$] (1)\citet{Furu08}; (2)\citet{Furu16}; (3)\citet{{Law07}}; (4)\citet{Grog11}; (5)\citet{Koek11}; (6)\citet{Ash13}; \\
    \footnotemark[$\dag$] The values show the total magnitude where the completeness of the source detection is 50\%.
  \end{tabnote}
\end{table*}
%---------- Table1 ----------%

%%%%%%%% 2.2 SED fitting %%%%%%%%
\subsection{ Sample selection and SED fitting } \label{S2s2ss}

 From the photometric sample, we select the objects satisfying all the following criteria.
\begin{itemize}
 \setlength{\itemsep}{-2pt}
\item[(1)] $i' \leq 26.0$
\item[(2)] Subaru/\textit{z'} or Subaru/updated-\textit{z'} $\geq$ 2$\,\sigma$
\item[(3)] UKIRT/\textit{J} or HST/F125W $\geq$ 2$\,\sigma$
\item[(4)] $B - R > 1.2$, $R - i' < 0.7$, and $B - R > 1.6(R - i') + 1.9$
\item[(5)] $3.5 \leq z_{phot} < 4.5$ with reduced $\chi^2 \leq 2$
\end{itemize}

 The criteria (1) is applied so as to select the galaxies bright enough to have small photometric errors, and the magnitude threshold corresponds to $S / N \gtrsim 11.5$. The criteria (2) and (3) are required to estimate the $\beta$ value accurately. Due to the stellar spikes and/or saturated pixels, some objects are not detected in the deeper Subaru/updated-\textit{z'} or HST/F125W imaging but detected in the shallower Subaru/\textit{z'} or UKIRT/\textit{J} imaging. Therefore, we use such the criteria (2) and (3). The criteria (4) is the \textit{BRi'}--LBG selection investigated by \citet{Ouch04} for the Subaru/Suprime-Cam filter set, and this criteria is intended to pick up star-forming galaxies at \textit{z} $\sim$ 4. In \citet{Ouch04}, the detectability of \textit{z} $\sim$ 4 galaxies and the contamination rate from low-\textit{z} galaxies are discussed, and leaders should refer to the reference for more details. After the criteria (1), (2), (3), and (4), the total number of objects is $\sim$ 2100. The criteria (5) is applied so as to select the reliable galaxies at \textit{z} $\sim$ 4. The reduced $\chi^{2}$ value is calculated for each galaxy as $\chi^2 / \mathrm{d.o.f}$, in which d.o.f $=$ (number of observed broad-band filters for each galaxy) $-$ (number of free parameters in the fitting). In the selection procedure, our concern is only the photometric redshift, and thus we adopt the number of free parameters $=$ 1. As a result, our catalog contains $\sim$ 1800 objects which are visually checked in order to avoid stellar spikes and/or saturated pixels.

 We write down the detail of $\sim$ 300 objects, which are excluded by the criteria (5). Among $\sim$ 300 objects, (i) $\sim$ 130 objects are the reduced $\chi^2 > 2$ objects at $3.5 \leq z_{phot} < 4.5$, (ii) $\sim$ 90 objects are the low-z interlopers at $z_{phot} < 3.0$, and (iii) $\sim$ 80 objects are the slightly lower/higher redshift objects at $3.0 \leq z_{phot} < 3.5$ or $4.5 \leq z_{phot} < 5.0$. First of all, it is reasonable that we exclude the objects in the sub-sample (ii) because they are the possible contamination in our study. On the other hand, the objects in the sub-sample (i) and (iii) are the potential LBGs at $3.5 \leq z_{phot} < 4.5$. We check the influence of the sub-sample (i) and (iii) on the UV slope $\beta$, the UV magnitude, and the other quantities, and consequently we confirm that the $\sim$ 300 objects excluded by the criteria (5) does not change our results and the criteria (5) is reasonable.

%---------- Table2 ----------%
\begin{table*}
  \tbl{Summary of parameters for SED fitting analysis}{%
  \begin{tabular}{ccc}
      \hline
      Parameter & \citet{Bruz03} & STARBURST99 \\
      \hline
      IMF & Chabrier & Kroupa \\
      SFH & Single Burst with finite time duration ($10^{7}\>\mathrm{Myr}$) & Instantaneous Burst \\
      & Continuous Constant & Continuous Constant \\
      & Exponentially Decline with $\tau = 0.1$, $2$, and $5\>\mathrm{Gyr}$ & \\
      Age & $5\>\mathrm{Myr}$--$15\>\mathrm{Gyr}$ & $2\>\mathrm{Myr}$--$150\>\mathrm{Myr}$ \\
      Metallicity ($Z$) & $0.02\,Z_{\solar}$, $0.2\,Z_{\solar}$, and $Z_{\solar}$ & $0.02\,Z_{\solar}$, $0.2\,Z_{\solar}$, $0.4\,Z_{\solar},$ and $Z_{\solar}$ \\
      Dust (Av) & $0.0$--$3.0$ & $0.0$--$3.0$ \\
      Dust extinction curve & \citet{Calz00} & \citet{Calz00} \\
      Redshift (z) & $0.0$--$6.0$ & $0.0$--$6.0$ \\
      Nebular continuum & Not included & Included and Not included \\ 
      \hline
  \end{tabular}}\label{tab2}
  \begin{tabnote}
  \end{tabnote}
\end{table*}
%---------- Table2 ----------%

 For the investigation of photometric redshift and stellar population, we use the Hyperz\footnote{$\langle$http://webast.ast.obs-mip.fr/hyperz/$\rangle$} photometric redshift code ver.1.1 \citep{Bolz00} with the \citet{Bruz03} templates\footnote{$\langle$http://www.bruzual.org/bc03/$\rangle$} (hereafter BC03) and the STARBURST99\footnote{$\langle$http://www.stsci.edu/science/starburst99/docs/default.htm$\rangle$} \citep{Leit99} templates (hereafter SB99). The BC03 templates are chosen as ''typical galaxy'' models and constructed from five different star formation histories (SFHs), thirty age values, and three metallicity values with the Chabrier Initial Mass Function (IMF). The SB99 templates are adopted as ''young star-forming galaxy'' models and constructed from two SFHs, thirty age values, four metallicity values, and two extreme nebular continuum cases with the Kroupa IMF. In the run of the Hyperz, the dust attenuation value, Av, is ranged from 0.0 to 3.0 with $\Delta \mathrm{Av} = 0.1$ assuming the \citet{Calz00} attenuation law for dust extinction curve. The details of the parameters are summarized in table \ref{tab2}. The motivation to use the BC03 and SB99 model templates simultaneously is that the BC03 template is not enough to describe young star-forming galaxies: A time step, which is critical to make spectra of young-age galaxies, in the SB99 computation is much smaller than that in the BC03 computation. Although the ages of the SB99 template set is very young, we consider that our template sets are optimal for fitting the spectrum of young star-forming galaxies.

 We note that we compare the photometry and the UV slope $\beta$ value used in this work with those of the published catalog. In the UDS-CANDELS field (the small area enclosed with the black dotted line in figure \ref{fig1}), a multi-wavelength photometry catalog has been published by the CANDELS collaborators \citep{Gala13}, and the catalog has the total flux of Subaru/\textit{BVRi'z'}, UKIRT/\textit{JHK}, HST/F125WF160W, and Spitzer/3.6$\>\micron$4.5$\>\micron$. Their photometry of the Subaru and UKIRT bands are consistent with ours if we take account of the difference in the measurement method. The UV slope $\beta$ value of our catalog is also comparable to that of the published catalog when applying the same measurement method for UV slope $\beta$ to both catalogs. However, the difference of the HST and Spitzer bands is slightly larger than expected. We consider that for the HST data the difference is attributed to our very large PSF correction factor applied to be matched with the Subaru images. For the Spitzer data the difference is due to the uncertainty in the aperture correction. We remark that our catalog tends to have the slightly bluer \textit{K} $-$ [3.6] and \textit{K} $-$ [4.5] colors than those of the published one.

%%%%%%%% 2.3 Example %%%%%%%%
\subsection{ Example results of SED fitting } \label{S2s3exs}

\begin{figure*}
  \begin{center}
    \includegraphics[width=140mm]{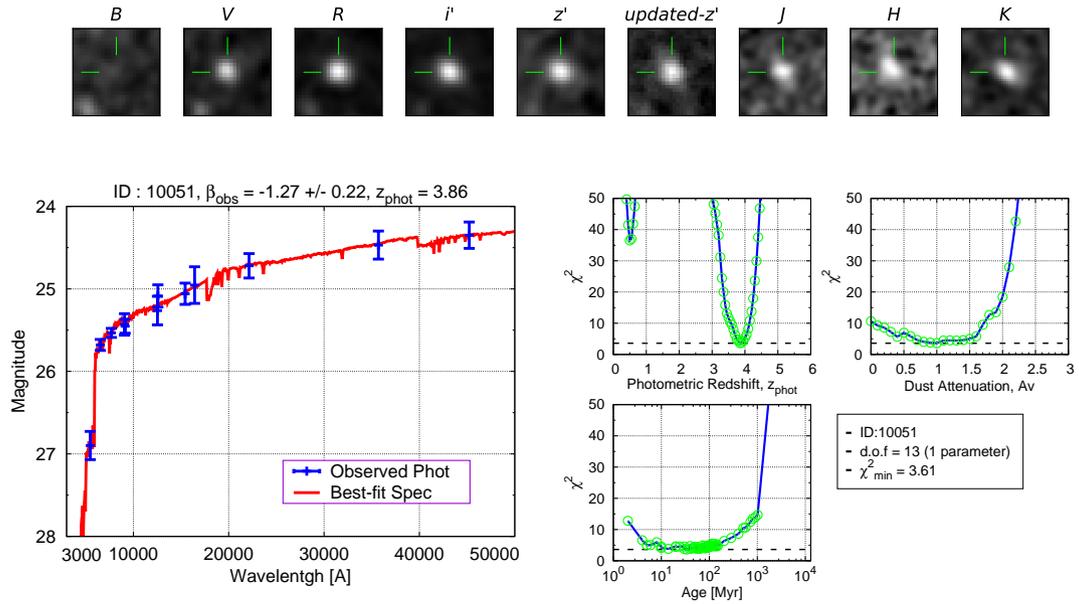}
  \end{center}
  \caption{ Example object for the red LBGs observed with Spitzer which has $\beta_{\mathrm{obs}} = -1.27$ and $M_{\mathrm{UV,obs}} = -20.38$. Top: The panels show the 5'' $\times$ 5'' stamps from Subaru/\textit{B} (left) to UKIRT/\textit{K} (right). The center of the images is the position of the detection. The green two straight lines in each stamp have 1'' length and are drawn at 1'' apart from the center. Bottom left: The blue plus points with the error bars show the measured aperture photometry with the 1\,$\sigma$ error from Subaru/\textit{B} (left) to Spitzer/3.6$\>\micron$ (right). The red solid line shows the best-fit SED model template. Bottom right: We show the $\chi^{2}$ map as a function of photometric redshift, dust attenuation, and age parameter. The horizontal black dashed line in each panel shows the minimum $\chi^{2}$ value. } \label{fig2}
\end{figure*}

\begin{figure*}
  \begin{center}
    \includegraphics[width=140mm]{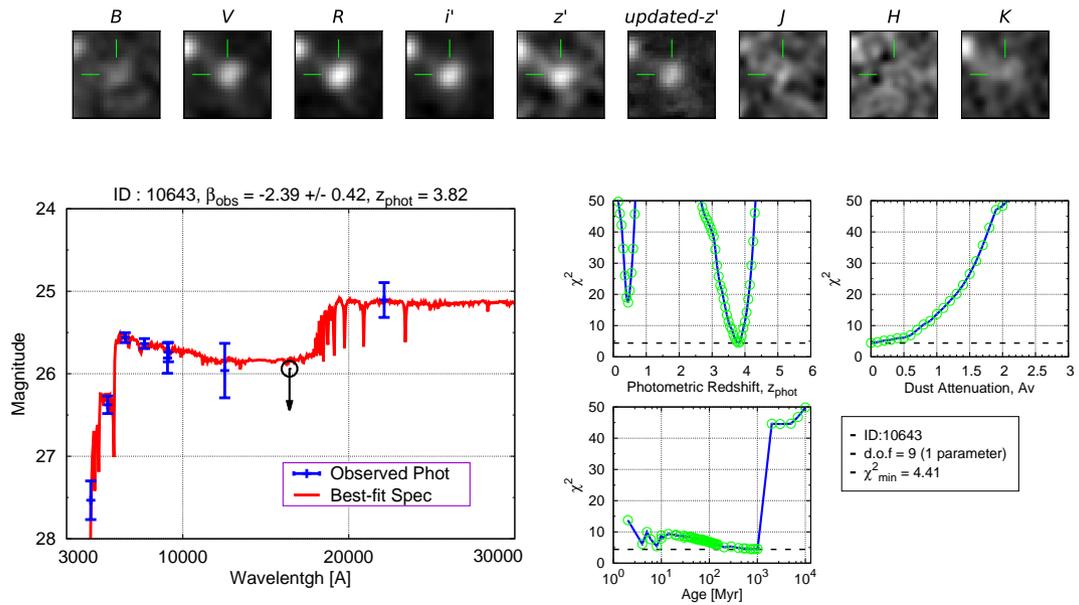}
  \end{center}
  \caption{ Same as figure \ref{fig2}, but the example object is the blue LBGs observed without Spitzer and it has $\beta_{\mathrm{obs}} = -2.39$ and $M_{\mathrm{UV,obs}} = -20.32$. } \label{fig3}
\end{figure*}

 We show two examples in figures \ref{fig2} and \ref{fig3} in order to show the validity of our SED fitting analysis. The first figure is the red LBG observed with Spitzer ($\beta_{\mathrm{obs}} = -1.27$ and $M_{\mathrm{UV,obs}} = -20.38$), and the second figure is the blue LBG observed without Spitzer ($\beta_{\mathrm{obs}} = -2.39$ and $M_{\mathrm{UV,obs}} = -20.32$). In the top nine panels of each figure, we show the stamps of the imaging data from Subaru/\textit{B} to UKIRT/\textit{K}. The center of the images represent the detected position, and the green two straight lines in each image with 1'' length are placed at 1'' from the detected position. In the bottom left panel, we show the observed photometry and best-fit SED. The blue plus points with the error bar represent the observed photometry and its uncertainty, and the red solid line represents the best-fit SED. The black open circle with the arrow represents the upper limit of the photometry at 2\,$\sigma$ level. In the bottom right three panels, we show the $\chi^{2}$ map of our SED fitting analysis on one-dimensional space. The vertical axis represents the $\chi^{2}$ value and the horizontal axis represents the photometric redshift, dust attenuation, and age. The horizontal black dashed line in each panel represents the minimum $\chi^{2}$ value. We emphasize that the two examples we show here are categorized as faintest objects, and most of the other objects are fitted better than the examples. Our SED fitting procedure performs well due to the deep imaging data of UKIRT/\textit{HK} which covers the wavelength of Balmer break for LBGs at \textit{z} $\sim$ 4.

%--------- Figure2 ---------%
%---------------------------%

%--------- Figure3 ---------%
%---------------------------%

%%%%%%%% 3 Measurement method %%%%%%%%
\section{ Measurement method of UV slope $\beta$ } \label{S3Mmuvb}

 According to the original definition of \citet{Calz94}, the UV slope $\beta$ value should be estimated from the Spectral Energy Distribution (SED) from 1250\,\AA\ to 2600\,\AA\ through the 10 fitting windows. However, spectroscopic observation generally requires a long exposure time for measuring the continuum flux from high-z galaxies due to their faint continuum. Furthermore, the number of objects which can be observed at a time is limited depending on slit configuration. Therefore, it is impractical to accurately measure the continuum flux from spectroscopic data for all the $\sim$ 1800 targets in our sample. Instead, we apply the simple power-law fitting to the broad-band photometry with the following functional form,
\begin{equation}
  M(\lambda_{x}) = -2.5 (\beta + 2) \log \lambda_{x} + Const
\label{eq1}
\end{equation}

\noindent where $\lambda_{x}$ is the effective wavelength of $x$th broad-band filter, $M$($\lambda_{x}$) is the measured magnitude of $x$th broad-band filter, and {\it Const} is a constant value. This method is suitable since the bias in the $\beta$ estimation is small (\cite{Fink12}; \cite{Rog13}). For the fitting, we conduct the least square fitting to the \textit{i'}, \textit{z'}, updated-\textit{z'}, and \textit{J} band filters which cover the rest-frame wavelength from $\sim$ 1500\,\AA\ to $\sim$ 2500\,\AA\ for the objects at \textit{z} $=$ 3.5--4.5. In our analysis, an uncertainty of the $\beta$ value represents the uncertainty of the least square fitting when taking account of the photometric error of the broad-band filters.

\begin{figure}
  \begin{center}
    \includegraphics[width=80mm]{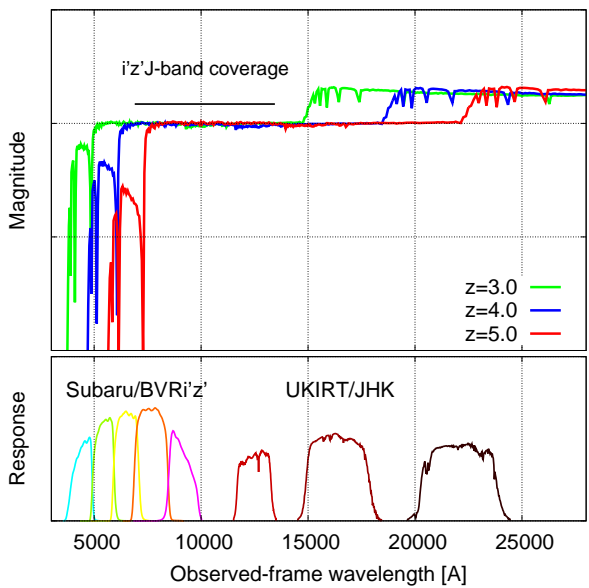}
  \end{center}
  \caption{ Model spectra and important broad-band filters. In the top panel, the green, blue, and red solid lines represent the model spectrum at \textit{z} $=$ 3, 4, and 5, respectively, which clearly show the Lyman and Balmer breaks. The length of the black horizontal line denotes the wavelength coverage of Subaru/\textit{i'}, Subaru/\textit{z'}, and UKIRT/\textit{J} band filters which are used for calculating the UV slope $\beta$. In the bottom panel, we show the Subaru/\textit{BVRi'z'} and UKIRT/\textit{JHK} band filters from left to right. For the sake of clarity, the filter responses of UKIRT/\textit{JHK} are multiplied by an arbitrary constant. } \label{fig4}
\end{figure}

 Although using the larger number of the photometry data points results in more accurate determination of the UV slope $\beta$ value, we need to select the optimal broad-band filters for the fitting so as to avoid strong spectral features. In the rest-frame UV and optical wavelength range, the redshifted Ly$\alpha$ ($\lambda$ $=$ 1216\,\AA) line (or Lyman break) and Balmer break ($\lambda$ $\sim$ 3600\,\AA) can be a contamination affecting the broad-band photometry. Figure \ref{fig4} illustrates the position of the Lyman and Balmer breaks in the observed-frame, and filter profiles of the optical and NIR broad-band. For the sake of clarity, we show three model spectra with clear Lyman and Balmer breaks at \textit{z} $=$ 3, 4, and 5, and we omit the filter profiles of Subaru/updated-\textit{z'}, HST/F125W, F160W, Spitzer/3.6$\>\micron$, and 4.5$\>\micron$ bands. This figure indicates that the \textit{R} ($\lambda$ $\sim$ 6000--7000\,\AA) and \textit{H} ($\lambda$ $\sim$ 15000-17000\,\AA) band filters are probably affected by Lyman or Blamer breaks, and the wavelength coverage from \textit{i'} to \textit{J} band (black horizontal line in upper panel of figure \ref{fig4}) is suitable for calculating the UV slope $\beta$ value of \textit{z} $\sim$ 4 LBGs. Consequently, we use the Subaru/\textit{i'}, \textit{z'}, updated-\textit{z'}, UKIRT/\textit{J}, and HST/F125W band filters.

 Another critical point is that the robustness of the $\beta$ measurement for the faint and blue galaxies is strongly influenced by the depth of the imaging data at longer wavelength in our analysis. As mentioned above, we select the $i' \leq 26.0$ LBGs, and we use the Subaru/\textit{i'}, \textit{z'}, updated-\textit{z'}, UKIRT/\textit{J}, and HST/F125W band filters for the $\beta$ measurement. Under the condition, for instance, the galaxies with $i' = 26.0$ and $\beta < -2.0$ have the larger photometric uncertainty in the \textit{z'}, updated-\textit{z'}, \textit{J}, and F125W band filters compared with the galaxies with $i' = 26.0$ and $\beta \geq -2.0$, and hence the bluer galaxies have the larger uncertainty in $\beta$. Moreover, the sample completeness of the extremely blue galaxies such as $\beta \sim -3.0$ is also influenced by the depth of the imaging data although the extremely blue objects are rare. From equation \ref{eq1}, the case of $\beta = -2.0$ is derived from $M(\lambda_{x}) = Const$, namely, all the broad-band photometry is same. For the appropriate $\beta$ measurement with the wide $\beta$ range, it is required that the magnitude threshold of \textit{i'} band is brighter than the $\sim$2--3$\sigma$ limiting magnitude of the \textit{z'}, updated-\textit{z'}, \textit{J}, and F125W bands. Our selection criteria described in section \ref{S2s2ss} is applied by taking the aspect into consideration. We use the conservative criteria and consider that our selection does not cause the strong bias in the $\beta$ distribution except for the extremely faint and blue objects. In order to quantify and discuss the influence, we estimate the recovery fraction in section \ref{S4s2RF}.

%--------- Figure4 ---------%
%--------- Figure4 ---------%

%%%%%%%% 4. Results %%%%%%%%
\section{Results} \label{S4Rslt}

%--------- Figure5 ---------%
\begin{figure}
  \begin{center}
    \includegraphics[width=80mm]{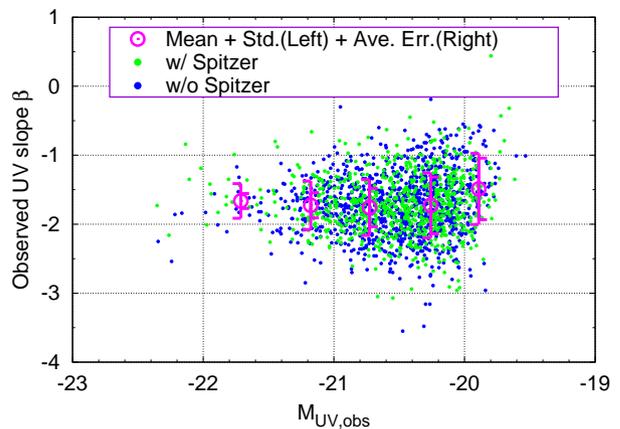}
  \end{center}
  \caption{ Observed UV slope $\beta$ vs. UV absolute magnitude at rest-frame 1500\,\AA\ ($\beta$--$M_{\mathrm{UV}}$ relation). The UV absolute magnitude is calculated by integrating the flux of the best-fit SED model template from rest-frame 1450\,\AA\ to 1550\,\AA. The green filled circles represent the individual objects in the area of the SEDS (Spitzer field) and the blue filled squares represent the individual objects out of the area. The magenta open circles with the error bar represent the mean UV slope $\beta$ value, standard deviation (left side), and mean error (right side) of the UV slope $\beta$ for each magnitude bin, and they are summarized in table \ref{tab3}. } \label{fig5}
\end{figure}
%--------- Figure5 ---------%

%%%%%%%% 4.1 Main Results %%%%%%%%
\subsection{Relation between $\beta$ and $M_{UV}$} \label{S4S1BM}

 Figure \ref{fig5} shows the obtained distribution of the UV slope $\beta$ as a function of the UV absolute magnitude at rest-frame 1500\,\AA\ ($\beta$--$M_{\mathrm{UV}}$ relation). The UV absolute magnitude for each object is calculated from the best-fit SED model template by integrating the flux from rest-frame 1450\,\AA\ to 1550\,\AA. The green filled circles show the objects in the area observed with Spitzer and the blue ones show the objects out of the area. There seems to be no notable systematic difference in the distribution of the sample with and without the Spitzer data, and the lack of the information about 3.6$\>\micron$ and 4.5$\>\micron$ does not influence the selection for \textit{z} $\sim$ 4 star-forming galaxies very much. In fact, we conduct the Kolmogorov--Smirnov test (K--S test) for the whole sample and some magnitude sub-samples. The null hypothesis of the K--S test is that the samples with (green) and without (blue) SEDS/Spitzer data are derived from the same distribution. The p-values of all the test are $>> 5\,\%$, and the null hypothesis is not rejected at 5\,$\%$ significance level. Therefore, there is no evidence that the information about 3.6$\>\micron$ and 4.5$\>\micron$ influences the $\beta$--$M_{\mathrm{UV}}$ relation.

 The magenta open circles with the error bar indicate the mean $\beta$ value and the mean $M_{\mathrm{UV}}$ value for each magnitude bin. The standard deviation of the $\beta$ distribution is indicated by the thick marks toward the left side, and the typical uncertainty in the $\beta$ value for the individual objects is shown by the thick marks toward the right side. For the mean values, we just apply the simple geometric mean without taking account of the individual uncertainty in $\beta$ for the individual objects. This is because the mean $\beta$ value can be biased toward positive values if we weight the $\beta$ values by the individual uncertainty of the objects, i.e., the uncertainty is not symmetric and becomes smaller toward positive $\beta$ values than the opposite. The mean $\beta$ value, standard deviation, typical uncertainty, and mean $M_{\mathrm{UV}}$ value for each bin are summarized in table \ref{tab3}. In addition, the other useful information such as the median values is also summarized.

%---------- Table3 ----------%
\begin{table*}
  \tbl{Summary of $\beta_{\mathrm{obs}}$--$M_{\mathrm{UV,obs}}$ Relation}{%
  \begin{tabular}{cccccccr}
      \hline
      $M_{\mathrm{UV}}$ bin & $M_{\mathrm{UV,mean}}$ & $M_{\mathrm{UV,median}}$ & $\beta_{\mathrm{mean}}$ & $\beta_{\mathrm{median}}$ & $\sigma_{\beta}$ & Mean Err. of $\beta$ & $N_{\mathrm{obj}}$ \\
      \hline
      $-$22.0 to $-$21.5 & $-$21.71 & $-$21.72 & $-$1.66 & $-$1.69 & 0.25 & 0.11 & 37 \\
      $-$21.5 to $-$21.0 & $-$21.18 & $-$21.15 & $-$1.73 & $-$1.71 & 0.35 & 0.18 & 204 \\
      $-$21.0 to $-$20.5 & $-$20.73 & $-$20.72 & $-$1.76 & $-$1.75 & 0.41 & 0.28 & 563 \\
      $-$20.5 to $-$20.0 & $-$20.26 & $-$20.26 & $-$1.73 & $-$1.74 & 0.47 & 0.41 & 861 \\
      $-$20.0 to $-$19.5 & $-$19.89 & $-$19.91 & $-$1.49 & $-$1.52 & 0.52 & 0.45 & 140 \\
      \hline
  \end{tabular}}\label{tab3}
  \begin{tabnote}
  \end{tabnote}
\end{table*}
%---------- Table3 ----------%

 The standard deviation is clearly larger than the typical uncertainty in the $\beta$ value except for the two of the most faint magnitude bin. Therefore, in the magnitude range of $M_{\mathrm{UV}} \lesssim -20.5$, the observed $\beta$ distribution is more scattered than the typical uncertainty and the observed scatter represents the variation of the stellar population and dust extinction among the sample. In the magnitude range of $M_{\mathrm{UV}} \gtrsim -20.5$, the typical uncertainty in $\beta$ becomes as large as the standard deviation, and particularly the objects with $M_{\mathrm{UV}} \gtrsim -20.0$ are not uniformly distributed. Although we does not add any criteria to our sample since our purpose is to investigate the bright LBGs with \textit{i'} $\leq$ 26.0, we check the sample completeness in section \ref{S4s2RF} and discuss the results with some $M_{\mathrm{UV}}$ sub-samples.

 The mean $\beta$ value seems not to decrease with the mean UV magnitude. In order to quantify the $\beta$--$M_{\mathrm{UV}}$ trend, we conduct the least-square linear fitting for our mean $\beta$ and $M_{\mathrm{UV}}$ values described in table \ref{tab3}. We use the four data points in $-22.0 \leq M_{\mathrm{UV}} \leq -20.0$, and the slope of the fitted linear equation becomes $-0.02\,\pm\,0.02$, which is nearly zero compared with the value in the previous works for the similar redshift, $-0.13\,\pm\,0.02$ \citep{Bouw14} and $-0.10\,\pm\,0.03$ \citep{Kurc14}. We note that our target is the relatively bright LBGs and the dynamic range of $M_{\mathrm{UV}}$ in our study is smaller than that in the previous works, $-22.0 \leq M_{\mathrm{UV}} \leq -15.5$ \citep{Bouw14} and $-21.0 \leq M_{\mathrm{UV}} \leq -15.0$ \citep{Kurc14}. When calculating the slope of the $\beta$--$M_{\mathrm{UV}}$ relation, we use \textit{standard deviation of the mean} as an uncertainty of each mean $\beta$ value. The uncertainty is calculated from the standard deviation divided by the number of galaxies in each bin ($= \sigma_{\beta} / N_{\mathrm{obj}}$) described in table \ref{tab3}, and thus the uncertainty of the slope of the $\beta$--$M_{\mathrm{UV}}$ relation is much smaller than the mean error of $\beta$.

\begin{figure}
  \begin{center}
    \includegraphics[width=80mm]{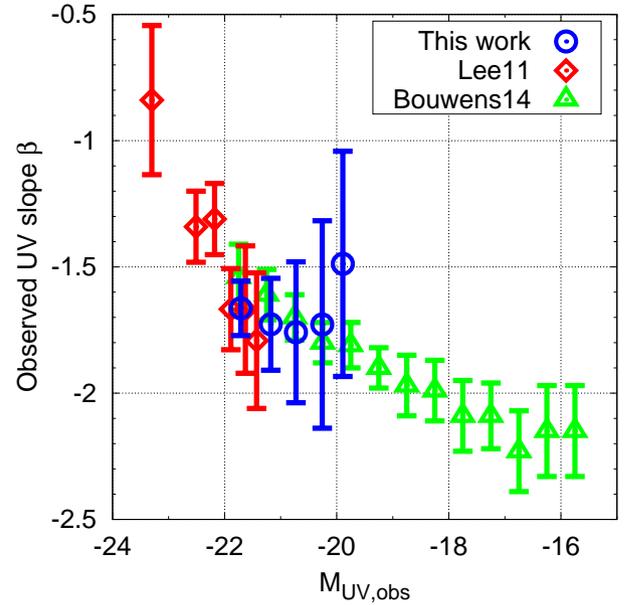}
  \end{center}
  \caption{ Comparison of this work with the previous studies at the similar redshift. For the sake of clarity, the vertical scale is changed from figure \ref{fig5}. The red open diamonds and green open triangle with the error bars show the results from \citet{LeeKS11} and \citet{Bouw14}, respectively. The blue open circles with the error bars show the result from our sample which is same as the magenta points in figure \ref{fig5}, but the error bars are the typical uncertainty. } \label{fig6}
\end{figure}

 In figure \ref{fig6}, we show the direct comparison of our result to the results from \textit{z} $\sim$ 4 super luminous stacked LBGs (\cite{LeeKS11}, red diamonds) and \textit{z} $\sim$ 4 faint LBGs (\cite{Bouw14}, green triangles). The red data points are calculated by us from the photometry described in \citet{LeeKS11} since all the $\beta$ values and its uncertainties are not described in their paper. The error bars denotes the typical uncertainty in the $\beta$ value. The error bars of \citet{Bouw14} show the sum of the random and systematic error described in their paper. The blue circles with the error bars show our result which is same as the magenta points in figure \ref{fig5} but the error bars are the typical uncertainty. In the magnitude range of $-22.0 \leq M_{\mathrm{UV}} \leq -20.0$, our results consistent with the previous works. Although the dynamic range of $M_{\mathrm{UV}}$ is smaller than the previous works, the luminous star-forming galaxies at \textit{z} $\sim$ 4, which are selected from the ground-based wide field images, seem to show the weaker $\beta$--$M_{\mathrm{UV}}$ relation over the magnitude range of $-22.0 \leq M_{\mathrm{UV}} \leq -20.0$.

%--------- Figure6 ---------%
%--------- Figure6 ---------%

 It seems that the distribution of the objects in figure \ref{fig5} is truncated and the shape of the distribution looks like a ``triangle''. This must be a result from either some physical constraint or some sample selection bias, or both. For example, a galaxy at \textit{z} $=$ 4.5 with the \textit{i'}-band magnitude \textit{i'} $= 26.0$ has $M_{UV} \lesssim -20.1$, and therefore the number of objects, which are selected from our selection criteria, decrease in the region around $M_{UV} \sim -20$ and $\beta \sim -2.5$. By using only our three data points in $-22.0 \leq M_{\mathrm{UV}} \leq -20.5$, the slope of the relation indeed becomes $-0.09\,\pm\,0.04$ which is quite similar to the value of the previous works. However, the number of the objects in the brightest (most left) bin is much less than the other bins (see column 8 in table \ref{tab3}), and the slope is almost estimated from only the two data points. We need to assess the incompleteness of the observed $\beta$ distribution in order to take account of our selection bias, which is discussed in the next section.

%%%%%%%% 4.2 Recovery Fraction %%%%%%%%
\subsection{Recovery Fraction} \label{S4s2RF}

 As shown in figure \ref{fig5}, the observed $\beta$ distribution is restricted in the ``triangle'' zone. It seems that there are three truncations, namely, (a) at the top left side, (b) at the bottom left side, and (c) at the bottom right side. In order to discuss the reason of those truncation and evaluate the validity of our results, we calculate the recovery fraction which is the number ratio of recovered objects to input objects by using Monte Carlo method.

 At first, we make a uniform input distribution on $\beta$--$M_{\mathrm{UV}}$ space for the quantitative discussion. For this purpose we consider the $8 \times 13 = 104$ grids with $\Delta \beta = 0.5$ and $\Delta M_{UV} = 0.25$, and we generate 300 mock galaxy spectra whose $\beta$ and $M_{UV}$ values place in each the small grid (so the total spectra are $104 \times 300 = 31,200$). The mock spectra are constructed from the BC03 or SB99 model templates which are the similar template sets as described in section \ref{S2s2ss}. All of the parameters such as SFH, dust attenuation value, age, metallicity, and source redshift are determined at random. We note that the range of the dust attenuation value and source redshift for the mock galaxies are different from the range described in section \ref{S2s2ss}, and they are $0.0 \leq {\rm Av} \leq 1.5$ and $3.5 \leq z_{s} \leq 4.5$. The number of the age step is also changed from 30 into 15. If a resultant spectrum does not place in the designated small grid, we repeat the procedure until the desired $\beta$ and $M_{\mathrm{UV}}$ values.

 Second, we calculate the apparent magnitude of the broad-band filters for each mock spectrum and we put the artificial galaxies on the real observed images from Subaru/\textit{B} to UKIRT/\textit{K} by using the IRAF mkobjects task. Since we check that the impact of Spitzer/3.6$\>\micron$ and 4.5$\>\micron$ is negligible for the $\beta$--$M_{\mathrm{UV}}$ relation in section \ref{S4S1BM}, we omit both the information in our simulation. The size and shape of the mock galaxies are also determined at random so that the size distribution of our simulated objects reproduces the observed size distribution.

 Finally these embedded mock galaxies are re-detected, re-measured, and re-compiled by the same manners described in section \ref{S2s2ss}. In the SED fitting procedure, however, we change the number of the age step from 30 to 15 in order to save the computer resource. After the compilation, we count the number of final recovered objects for each small grid and calculate the number ratio of recovered to input objects. The final result includes the impact from the image quality, the magnitude criteria, the LBG selection, and the photo-\textit{z} selection. Note that the prepared objects are restricted by only the rest-frame {\it UV} information and thus the rest-frame {\it optical} information such as Balmer break is purely determined at random.

%--------- Figure7 ---------%
\begin{figure}
  \begin{center}
    \includegraphics[width=80mm]{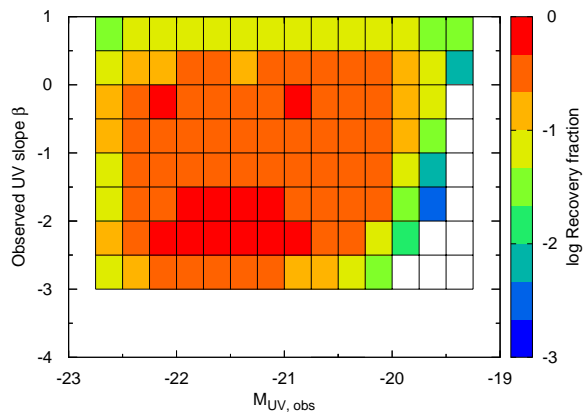}
  \end{center}
\caption{ Recovery fraction which is the number ratio of recovered to input objects on $\beta$--$M_{\mathrm{UV}}$ space. The black grid lines represent the area where we prepare the input objects uniformly throughout the $\beta$--$M_{\mathrm{UV}}$ space and a total of 31,200 mock galaxies is distributed. The colored area represents the detected area where we can find recovered objects. The white colored area represents the non-detection area where we cannot find any recovered objects. }\label{fig7}
\end{figure}
%--------- Figure7 ---------%

 Figure \ref{fig7} shows the final recovery fraction map by the color-coded area. The vertical axis is the UV slope $\beta$ and the horizontal axis is the absolute magnitude at rest-frame 1500\,\AA. Although the UV absolute magnitude for input objects is given as total magnitude, the UV absolute magnitude for recovered objects is calculated from 2''-diameter aperture photometry. Therefore we convert the total magnitude of input objects to the 2'' aperture magnitude by the aperture correction: $M_{\mathrm{UV,aperture}} = M_{\mathrm{UV,total}} + 0.352$. The black lattice lines indicate each area where $\sim$ 300 mock galaxies (or input objects) are distributed except for both the faintest and brightest magnitude bins where $\sim$ 150 mock galaxies are distributed. The white area represents the non-detection area which means that there are no recovered objects.

 We find that the relative value of the recovery fraction is roughly homogeneous over the area where the observed objects are distributed ($-2.5 \lesssim \beta \lesssim -0.5$ and $-22.0 \lesssim M_{\mathrm{UV}} \lesssim -20.0$) and does not depend on the UV magnitude except for the area around the truncation (c). The rough homogeneity of the relative value indicates that the observed $\beta$--$M_{\mathrm{UV}}$ distribution is not strongly biased at least over the area of $-2.5 \leq \beta \leq -0.5$ and $-22.0 \leq M_{\mathrm{UV}} \leq -20.0$, and our measurement for the $\beta$--$M_{\mathrm{UV}}$ relation described in section \ref{S4S1BM} is reasonable. In other words, we find no evidence that the truncation (a) and (b) are artificial, and they must be caused by some other reasons. On the other hand, this figure also shows that the truncation (c) is artificially caused by our sample selection. Our simulation indicates that the truncation (c) is attributed to the selection criteria of the detection in Subaru/\textit{z} or Subaru/updated-\textit{z'}, and UKIRT/\textit{J} or HST/F125W.

 The figure also indicates that the recovery fraction locally peaks around $\beta \sim -2.0$ and there are some fluctuated peaks at $\beta \sim -0.25$. Qualitatively, prominent spectral features can be easily identified by the SED fitting procedure, and then the recovery fraction may become higher than the other $\beta$ values. Since the Lyman Break technique prefers to select blue galaxies such as $\beta \sim -2.0$, our simulation indeed reflects our sample selection rather than the assumption about input objects. In addition, red galaxies such as $\beta \sim -0.25$ have clear Balmer Break if their red color is due to the aged stellar population. In our simulation the rest-frame optical information is determined at random, and hence the too many input galaxies, which have the prominent Balmer Break, are probably generated to have $\beta \sim -0.25$. In conclusion, the inhomogeneity of the recovery fraction seen in figure \ref{fig7} is due to our sample selection criteria adopting the Lyman Break technique and photo-\textit{z} estimation.

 The truncation (a) and (b) can be interpreted as follows: The observed number of LBGs decreases toward the brighter UV magnitude and the average $\beta$ value converges in $\beta \sim -1.7$. The decrease of LBGs along with the UV magnitude is explained by the drop of UV Luminosity Function since the characteristic luminosity of \textit{z} $\sim$ 4 LBGs is $M_{\mathrm{UV}}^{*} = -21.14$ \citep{Yoshi06}. However, it is not clear why the average UV slope $\beta$ value converges in $\beta \sim -1.7$. Qualitatively the galaxies with $\beta \gtrsim -1.7$ should contain a large amount of dust and their UV magnitude becomes fainter due to the dust obscuration. Therefore the red and bright galaxies are a rare or almost impossible population, and it causes the truncation (a). On the other hand, the galaxies with $\beta \lesssim -1.7$ contain a less amount of dust and the galaxies can remain bright UV magnitude. As we cannot find the blue and bright galaxies from figure \ref{fig5}, such the objects are indeed a rare population in the observational data. 

 In summary, we conclude that the truncation (a) and (b) are not only caused by our sample selection and are most likely caused by some physical requirements, and the truncation (c) is clearly caused by our sample selection. In order to understand what make the blue and bright galaxies rare and to reveal the reason of the truncation (b), we discuss the underlying stellar population of LBGs for our sample in section \ref{S5Dis}.

%%%%%%%% 5. Discussions %%%%%%%%
\section{Discussion}  \label{S5Dis}

%%%%%%%% 5.1 Intrinsic Beta & Muv %%%%%%%%
\subsection{Relation between Intrinsic $\beta$ and Intrinsic $M_{\mathrm{UV}}$} \label{S5s1IBM}

 We here estimate the dust-corrected $\beta$ (hereafter we call it {\it intrinsic} UV slope, $\beta_{\mathrm{int}}$) and the dust-corrected $M_{\mathrm{UV}}$ (hereafter we call it {\it intrinsic} UV absolute magnitude, $M_{\mathrm{UV,int}}$). In section \ref{S4s2RF}, our simulation indicates that the observed distribution on $\beta$--$M_{\mathrm{UV}}$ space is caused by some physical reasons. Both observed $\beta$ and $M_{\mathrm{UV}}$ value strongly depend on the dust attenuation value, and hence it is helpful to investigate the $\beta$--$M_{\mathrm{UV}}$ distribution before the dust reddening. In our discussion, we assume that the reasonable best-fit physical quantities are estimated from the SED fitting analysis in which the observed photometry covers the wavelength range between rest-frame $\sim$ 900\,\AA\ and $\sim$ 4400\,\AA\ (or $\sim$ 9000\,\AA\ in part) for \textit{z} $\sim$ 4 objects.

%--------- Figure8 ---------%
\begin{figure*}
  \begin{center}
    \includegraphics[width=140mm]{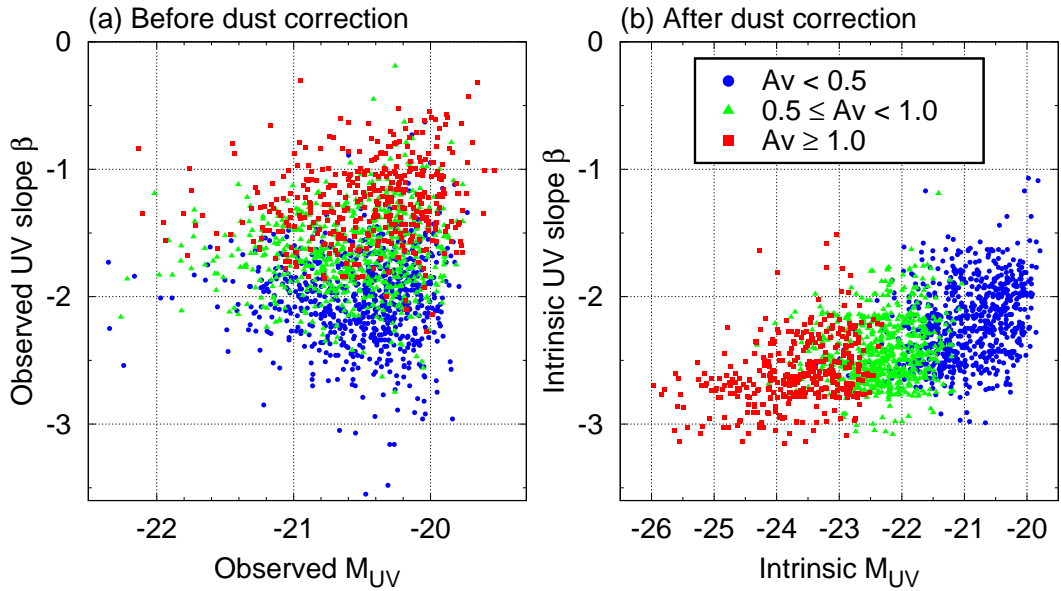}
  \end{center}
\caption{ Comparison of the distribution for (a) observed UV slope $\beta$ vs. observed absolute magnitude at rest-frame 1500\,\AA\ (left, same as figure \ref{fig4}) and (b) \textit{intrinsic} UV slope $\beta$ vs. \textit{intrinsic} absolute magnitude (right). In both panel, the best-fit dust attenuation values for the individual objects are expressed by the blue, green, and red color-coding, which indicate Av $<$ 0.5, 0.5 $\leq$ Av $<$ 1.0, and Av $\geq$ 1.0, respectively. }\label{fig8}
\end{figure*}
%--------- Figure8 ---------%

 We calculate the intrinsic UV slope by equation \ref{eq1} for the intrinsic magnitude of Subaru/\textit{i'}, Subaru/updated-\textit{z'}, and UKIRT/\textit{J} band filters. For estimating the intrinsic magnitude, we convolve the intrinsic SED, which is reproduced with the best-fit physical quantities without any dust extinction, with the three broad-band filters. We note that the intrinsic UV slope depends on the prepared model templates in the SED fitting (i.e., SFH, age, and metallicity) and has discrete values in out discussion.

 Figure \ref{fig8} shows the conversion from observed to intrinsic value for $\beta$ and $M_{\mathrm{UV}}$. The left panel (a) of figure \ref{fig8} shows the \textit{observed} $\beta$ as a function of the \textit{observed} $M_{\mathrm{UV}}$ ($\beta_{\mathrm{obs}}$--$M_{\mathrm{UV,obs}}$ relation, same as figure \ref{fig4}). The right panel (b) of figure \ref{fig8} shows the \textit{intrinsic} $\beta$ as a function of the \textit{intrinsic} $M_{\mathrm{UV}}$ ($\beta_{\mathrm{int}}$--$M_{\mathrm{UV,int}}$ relation). The blue filled circles, green filled triangles, and red filled squares represent individual objects with the best-fit dust attenuation value of Av $<$ 0.5, 0.5 $\leq$ Av $<$ 1.0, and Av $\geq$ 1.0, respectively. In the panel (a), we confirm that the objects with the higher dust attenuation value are distributed at the upper area where the $\beta_{\mathrm{obs}}$ value becomes redder. This trend is natural and is not inconsistent with the previous studies reported as the relation between the $\beta_{\mathrm{obs}}$ and dust attenuation value (IRX--$\beta$ relation: e.g., \cite{Calz94}; \cite{Meur99}; \cite{Take12}). In the panel (b), due to the large dust correction, the objects with the higher dust attenuation value and the redder $\beta_{\mathrm{obs}}$ value tend to be distributed at the bottom left area where the $\beta_{\mathrm{int}}$ and $M_{\mathrm{UV,int}}$ value becomes bluer and brighter. Moreover, the trend of the distribution is different from the panel (a), namely, the slope of the $\beta$--$M_{\mathrm{UV}}$ relation is nearly constant or positive. We discuss this distribution for different sub-samples in the following.

\begin{figure*}
  \begin{center}
    \includegraphics[width=70mm]{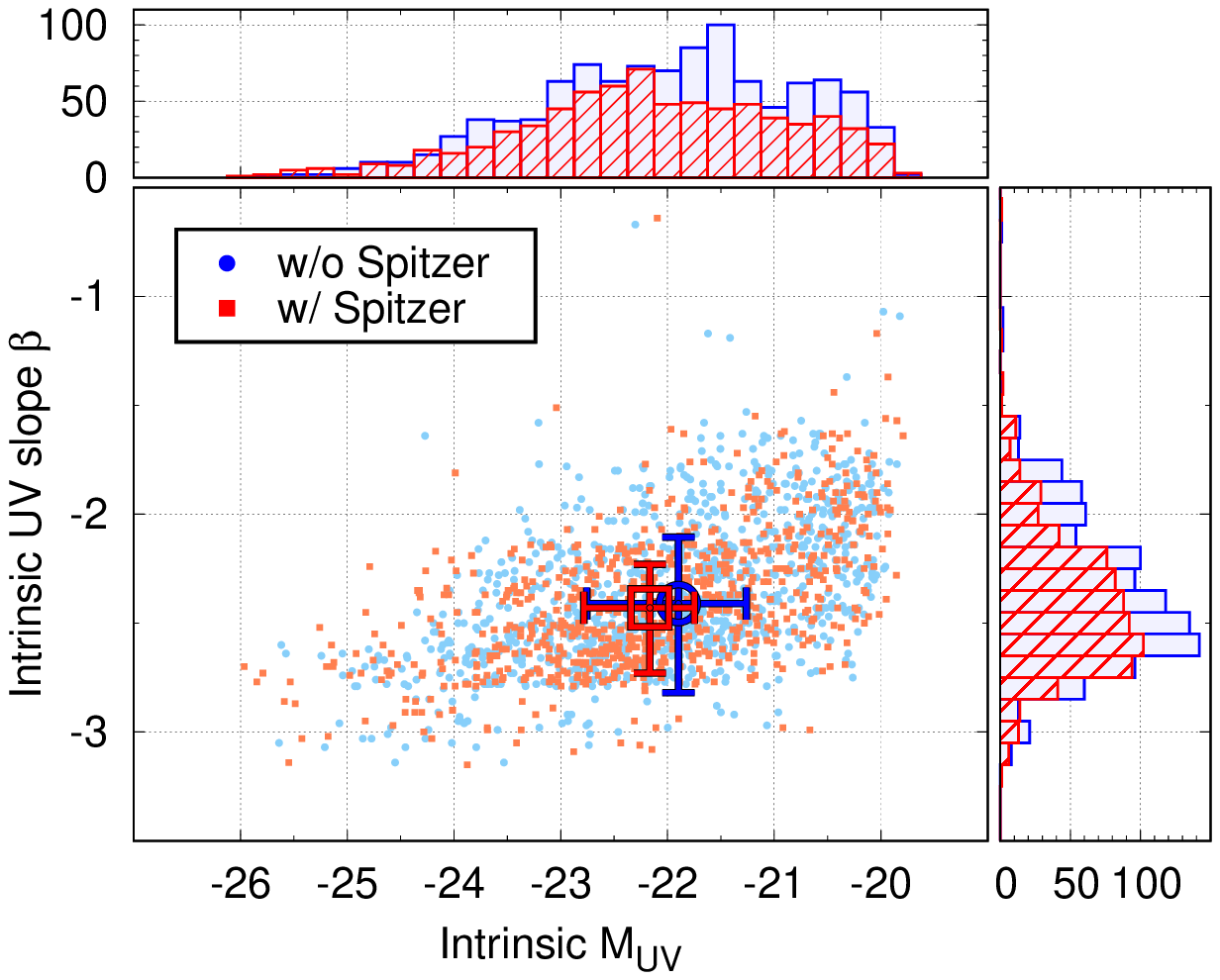}
    \includegraphics[width=70mm]{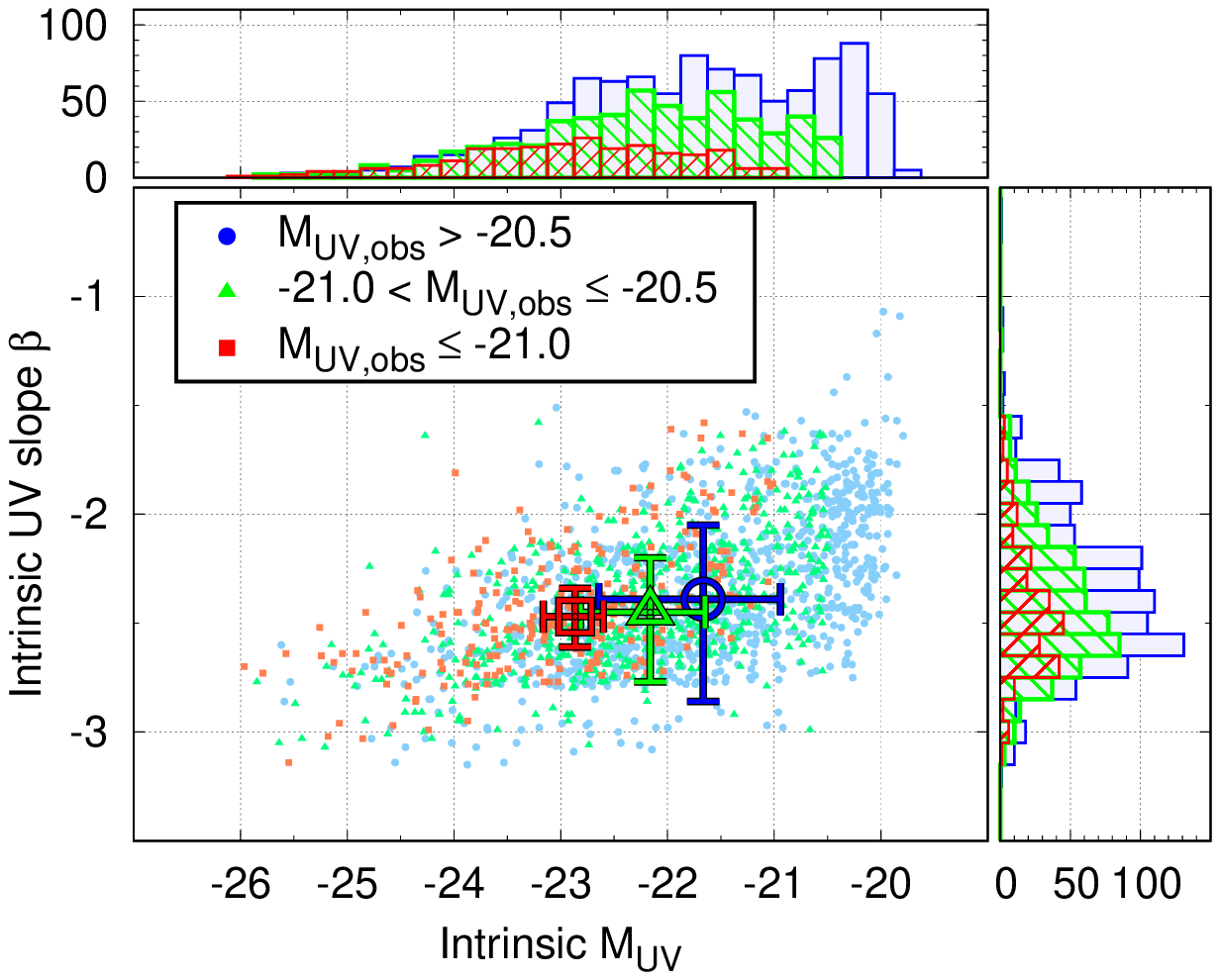}    
  \end{center}
\caption{ The $\beta_{\mathrm{int}}$--$M_{\mathrm{UV,int}}$ distribution of sub-samples classified according to with and without SEDS/Spitzer (left) and $M_{\mathrm{UV,obs}}$ (right). The open circle, triangle, and square with the error bars in each panel represent the median value and the median uncertainty for each sub-sample. The uncertainty in $M_{\mathrm{UV,int}}$ and $\beta_{\mathrm{int}}$ is estimated from the uncertainty in Av. }\label{fig910}
\end{figure*}

 Both two panels in figure \ref{fig910} shows the same $\beta_{\mathrm{int}}$--$M_{\mathrm{UV,int}}$ distribution but the color-coding represent the different sub-samples classified according to with and without SEDS/Spitzer (left) and $M_{\mathrm{UV,obs}}$ (right). In the left panel, the red filled squares and blue filled circles indicate the objects with and without SEDS/Spitzer data, respectively. In the right panel, the blue filled circles, green filled triangles, and red filed squares indicate the objects with $M_{\mathrm{UV,obs}} > -20.5$, $-20.5 \geq M_{\mathrm{UV,obs}} > -21.0$, and $M_{\mathrm{UV,obs}} < -21.0$, respectively. The large open circle, triangle, and square with the error bars in each panel represent the median value and the median uncertainty for each sub-sample. We calculate the dust attenuation value at $\chi^{2}_{min} + 1$ for the individual objects as the uncertainty of the dust attenuation, and then the uncertainty in Av is converted into the uncertainty in $\beta_{\mathrm{int}}$ and $M_{\mathrm{UV,int}}$. Therefore, the error bars in figure \ref{fig910} denote the uncertainty in Av. We also show the histogram of $\beta_{\mathrm{int}}$ and $M_{\mathrm{UV,int}}$ for each sub-sample.

%--------- Figure9,10 ---------%
%--------- Figure9,10 ---------%

 From figure \ref{fig910}, there seems not to be systematic difference in the $\beta_{\mathrm{int}}$ distribution. From the left panel, we find that the information of the Spitzer data does not influence the estimation of the $\beta_{\mathrm{int}}$ value although the uncertainty in $\beta_{\mathrm{int}}$ and $M_{\mathrm{UV,int}}$ for the sub-sample with Spitzer tends to be smaller than that for the sub-sample without Spitzer. From the right panel, we find that the $\beta_{\mathrm{int}}$--$M_{\mathrm{UV,int}}$ distribution of each $M_{\mathrm{UV,obs}}$ sub-sample is almost parallel to each other. When calculating the slope of the $\beta_{\mathrm{int}}$--$M_{\mathrm{UV,int}}$ relation for each sub-sample by the same manner described in section \ref{S4S1BM}, we obtain the value of $0.12 \pm 0.01$ ($M_{\mathrm{UV,obs}} > -20.5$), $0.14 \pm 0.01$ ($-20.5 \geq M_{\mathrm{UV,obs}} > -21.0$), and $0.16 \pm 0.02$ ($M_{\mathrm{UV,obs}} < -21.0$). It means that the variation of the $M_{\mathrm{UV,obs}}$ value does not significantly affect the shape of the $\beta_{\mathrm{int}}$--$M_{\mathrm{UV,int}}$ distribution. Although the faint objects such as the objects with $M_{\mathrm{UV,obs}} > -20.5$ have the $\beta_{\mathrm{obs}}$ value with the large uncertainty, the $\beta_{\mathrm{int}}$ values are reasonably well evaluated after the SED fitting using the all photometric data points.

%--------- Figure11 ---------%
\begin{figure*}
  \begin{center}
    \includegraphics[width=140mm]{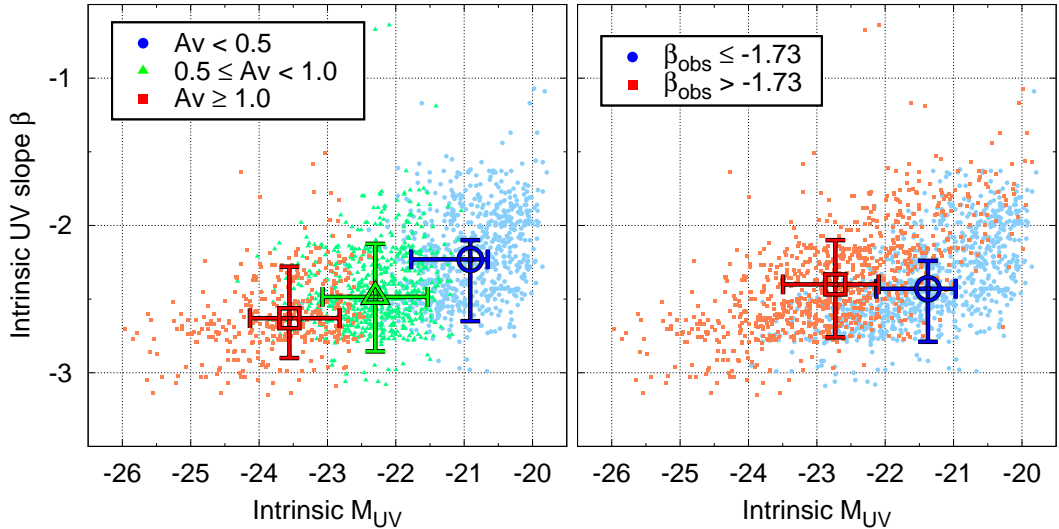}
  \end{center}
\caption{ The $\beta_{\mathrm{int}}$--$M_{\mathrm{UV,int}}$ distribution of sub-samples classified according to Av and $\beta_{\mathrm{obs}}$. The open circle, triangle, and square with the error bars in each panel represent the median value and the median uncertainty for each sub-sample. The uncertainty in $M_{\mathrm{UV,int}}$ and $\beta_{\mathrm{int}}$ is estimated from the uncertainty in Av. }\label{fig11}
\end{figure*}
%--------- Figure11 ---------%

 Figure \ref{fig11} shows the same $\beta_{\mathrm{int}}$--$M_{\mathrm{UV,int}}$ distribution as figure \ref{fig910} but the color-coding represent the sub-samples classified according to Av and $\beta_{\mathrm{obs}}$. In the left panel, the blue filled circles, green filled triangles, and red filled squares indicate the objects with Av $<$ 0.5, 0.5 $\leq$ Av $<$ 1.0, and Av $\geq$ 1.0, respectively. In the right panel, the blue filled circles and red filled squares indicate the objects with $\beta_{\mathrm{obs}} \leq -1.73$ and $\beta_{\mathrm{obs}} > -1.73$, respectively. $\beta_{\mathrm{obs}} = -1.73$ represents the median $\beta_{\mathrm{obs}}$ value for our whole sample. The open circle, triangle, and square with the error bars in each panel represent the median value and the median uncertainty for each sub-sample. As mentioned above, the uncertainty in $\beta_{\mathrm{int}}$ and $M_{\mathrm{UV,int}}$ is estimated from the uncertainty in Av.

 Figure \ref{fig11}, interestingly, shows that the objects which are dusty and redder in the observed $\beta$ tend to be bluer in the intrinsic $\beta$ and brighter in the intrinsic $M_{\mathrm{UV}}$. In addition, the intrinsic $\beta$ value slightly increases with the intrinsic $M_{\mathrm{UV}}$ value and the trend is opposite of that of the $\beta_{\mathrm{obs}}$--$M_{\mathrm{UV,obs}}$ relation. Our result can be interpreted as follows. The more intense ongoing star-forming galaxies, whose {\it intrinsic} $\beta$ and $M_{\mathrm{UV}}$ value are bluer and brighter, generate and/or contain a large amount of dust, and the {\it observed} $\beta$ and $M_{\mathrm{UV}}$ value result in a redder and fainter value due to the dust attenuation. Then, the nearly constant $\beta_{\mathrm{obs}}$--$M_{\mathrm{UV,obs}}$ distribution observed in our analysis is formed by the galaxies which have a blue $\beta_{\mathrm{int}}$ and bright $M_{\mathrm{UV,int}}$ value because they are distributed at the area of a red $\beta_{\mathrm{obs}}$ and faint $M_{\mathrm{UV,obs}}$ value.

 According to our SED fitting analysis, a young-age stellar population is responsible for the bluest $\beta_{\mathrm{int}}$ value. In other words, there are some young-age galaxies with the bluest $\beta_{\mathrm{int}}$ value in the brightest $M_{\mathrm{UV,int}}$ range, but there are no intermediate-age and old-age galaxies with the bluest $\beta_{\mathrm{int}}$ value in the brightest $M_{\mathrm{UV,int}}$ range. This is not surprising because the intrinsic UV luminosity is expected to be sensitive to the age of the stellar population. It is hard to sustain a very high star formation rate with the intermediate and long time duration due to rapid gas depletion. The UV luminosity is dominated by the stars at ''turn-off point'' on Hertzsprung--Russell Diagram which is an age indicator of the young stellar population. We, however, emphasize that other parameters such as metallicity and/or IMF can also explain the reason of the bluest $\beta_{\mathrm{int}}$ value. Indeed some literature argue that dusty star-forming galaxies have a ''top-heavy'' IMF although the discussion still continues (\cite{Bau05}, \cite{Tacc08}, \cite{Bas10}). Under the top-heavy IMF environment, hot and massive stars can be formed more and more, and the bluer $\beta_{\mathrm{int}}$ value is easily produced. Otherwise, among the galaxies with the bluest $\beta_{\mathrm{int}}$ value, there may be a post-primordial starburst which is dominated by extremely metal-poor (or PopI\hspace{-.1em}I\hspace{-.1em}I) stars.

\begin{figure}
  \begin{center}
    \includegraphics[width=80mm]{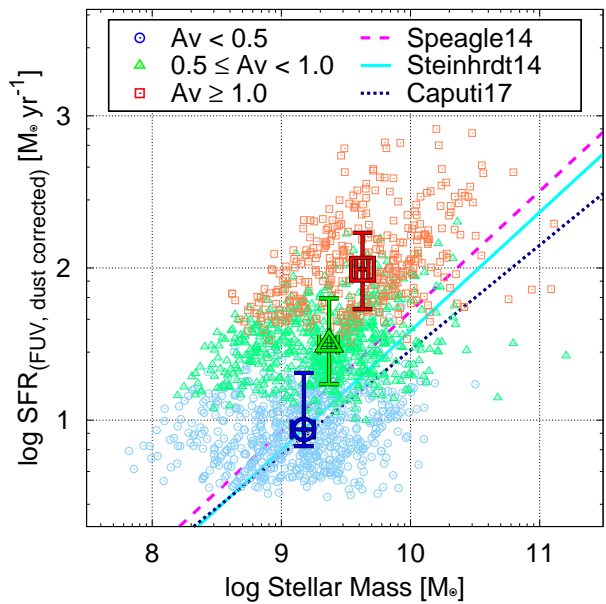}
  \end{center}
  \caption{ Dust-corrected SFR vs. stellar mass estimated from the BC03 model templates. The SFR is estimated from the luminosity at the \textit{GALEX}/FUV filter by using the \citet{Hao11} calibration. The stellar mass is calculated multiplying the normalization factor to the output of the BC03 model. The color-coding represents the best-fit dust attenuation value. We also draw the SFR--$\mathrm{M_{*}}$ relation at \textit{z} $\sim$ 4 reported by the previous works (\cite{Spe14}, \cite{Ste14}, \cite{Cap17}). } \label{fig12}
\end{figure}

 In order to investigate the star formation activity, we plot the Star Formation Rate (SFR) of the individual objects as a function of their stellar mass in figure \ref{fig12}. For the estimation of SFR, we convolve the best-fit template with the GALEX/FUV filter response curve and use the calibration for FUV luminosity (\cite{Hao11}; \cite{Kenn12}). For estimating the stellar mass, we multiply the best-fit normalization factor to the output from the BC03 model template. Figure \ref{fig12} shows the dust-corrected SFR as a function of the stellar mass, and the blue circle, green triangle, and red squares represent the individual objects with the dust attenuation value of Av $<$ 0.5, 0.5 $\leq$ Av $<$ 1.0, and Av $\geq$ 1.0, respectively. The large open circle, triangle, and square with the error bars show the median value and median uncertainty of each sub-sample. The uncertainty in SFR is estimated from the uncertainty in Av, and the uncertainty in stellar mass is estimated from the photometric uncertainty of \textit{K} band since the estimation of the stellar mass is almost determined by the \textit{K} band photometry. We also plot the previous results of the SFR--$\mathrm{M_{*}}$ relation called as main-sequence of star forming galaxies at \textit{z} $\sim$ 4 (\cite{Spe14}, \cite{Ste14}, \cite{Cap17}).

%--------- Figure12 ---------%
%--------- Figure12 ---------%

 The figure shows that the most intense star-forming galaxies in our sample have SFR $\gtrsim$ a few $\times\ 10^{2}\,\mathrm{M_{\solar}}\>\mathrm{yr^{-1}}$, and they are the objects with Av $\geq$ 1.0. Since most of the objects with Av $\geq$ 1.0 have $\beta_{\mathrm{obs}} > -1.73$ and $\beta_{\mathrm{int}} \leq -2.42$ ($\beta_{\mathrm{int}} = -2.42$ is the median $\beta_{\mathrm{int}}$ value for our whole sample) from figure \ref{fig11}, our analysis indicates that the highly dust-attenuated and intense star-forming galaxies at \textit{z} $\sim$ 4 tend to have $\beta_{\mathrm{obs}} > -1.73$ and $\beta_{\mathrm{int}} \leq -2.42$. When comparing our result with the previous works, our median value of Av $<$ 0.5 and 0.5 $\leq$ Av $<$ 1.0 sub-sample is consistent with the relation from the previous works although the distribution of our sample is significantly scattered. The sub-sample with Av $\geq$ 1.0 tends to be distributed above the relation from the previous works, and the deviation of the median value from the relation is larger than the uncertainty in SFR. Because the galaxies distributed above the star formation main sequence are classified as starburst phase (e.g., \cite{Cap17}; \cite{Bisi18}), we consider that the objects with Av $\geq$ 1.0 are indeed in the starburst phase and our results is not inconsistent with the previous works. In conclusion, we find some highly dust-attenuated (Av $\geq$ 1.0) and intense star-forming (SFR $\gtrsim$ a few $\times\ 10^{2}\,\mathrm{M_{\solar}}\>\mathrm{yr^{-1}}$) galaxies at \textit{z} $\sim$ 4 which have $\beta_{\mathrm{obs}} > -1.73$ and $\beta_{\mathrm{int}} \leq -2.42$.

 Finally, we consider a simple case in which the $\beta_{\mathrm{int}}$--$M_{\mathrm{UV,int}}$ trend monotonically continues in the fainter magnitude range. According to \citet{Bouw14}, the $\beta_{\mathrm{obs}}$ value becomes bluer when the $M_{\mathrm{UV,obs}}$ value becomes fainter, but the slope of the $\beta_{\mathrm{obs}}$--$M_{\mathrm{UV,obs}}$ relation becomes flatter in $M_{\mathrm{UV,obs}} \gtrsim -19.0$. In order to establish both the observed and intrinsic $\beta$--$M_{\mathrm{UV}}$ relation without contradiction, it is expected that the $\beta_{\mathrm{int}}$ value becomes redder and converges to the certain $\beta$ value toward the fainter magnitude range. When we extrapolate the $\beta_{\mathrm{int}}$--$M_{\mathrm{UV,int}}$ relation faintward below our sample magnitude limit, we will find the intersection point of the observed and intrinsic $\beta$--$M_{\mathrm{UV}}$ relation. Since the dust attenuation value becomes smaller toward the fainter magnitude range along the $\beta_{\mathrm{int}}$--$M_{\mathrm{UV,int}}$ relation, the intersection point (or convergence point) will represent the position of the appearance of nearly dust-free population. Our $\beta_{\mathrm{int}}$--$M_{\mathrm{UV,int}}$ relation shows $\beta_{\mathrm{int}} = 0.61 + 0.14 M_{\mathrm{UV,int}}$ by the same manner described in section \ref{S4S1BM}, and the $\beta_{\mathrm{obs}}$--$M_{\mathrm{UV,obs}}$ relation from \citet{Bouw14} shows $\beta_{\mathrm{obs}} = -4.39 - 0.13 M_{\mathrm{UV,obs}}$ in $M_{\mathrm{UV,obs}} \leq -18.8$. As a result, both relations intersect at $M_{\mathrm{UV}} = -18.9$ and $\beta = -1.94$ and its point corresponds to the break point of $\beta_{\mathrm{obs}}$--$M_{\mathrm{UV,obs}}$ relation at $M_{\mathrm{UV}} = -18.8$ and $\beta = -1.95$ reported by \citet{Bouw14}. Therefore the transition of the $\beta_{\mathrm{obs}}$-$M_{\mathrm{UV,obs}}$ relation around $M_{\mathrm{UV}} \sim -18.8$ indicates that we really see the almost dust-free population in $M_{\mathrm{UV}} > -18.8$, and the apparently bluest star-forming galaxies at \textit{z} $\sim$ 4 distribute around $\beta \sim -2.0$.

%%%%%%%% 5.2 validate result %%%%%%%%
\subsection{ Case of fixed star formation history and SMC attenuation law } \label{S5s2fhsmc}

In this section and the following sections (section \ref{S5s3zjkd} and \ref{S5s4iasfg}), we verify our results by using different and somewhat independent ways. We emphasize that these verification is intended not only to check the results from our SED fitting analysis but also to strengthen our suggestion, i.e., we find the dusty and on-going active star-forming galaxies at \textit{z} $\sim$ 4.

\begin{figure*}
  \begin{center}
    \includegraphics[width=140mm]{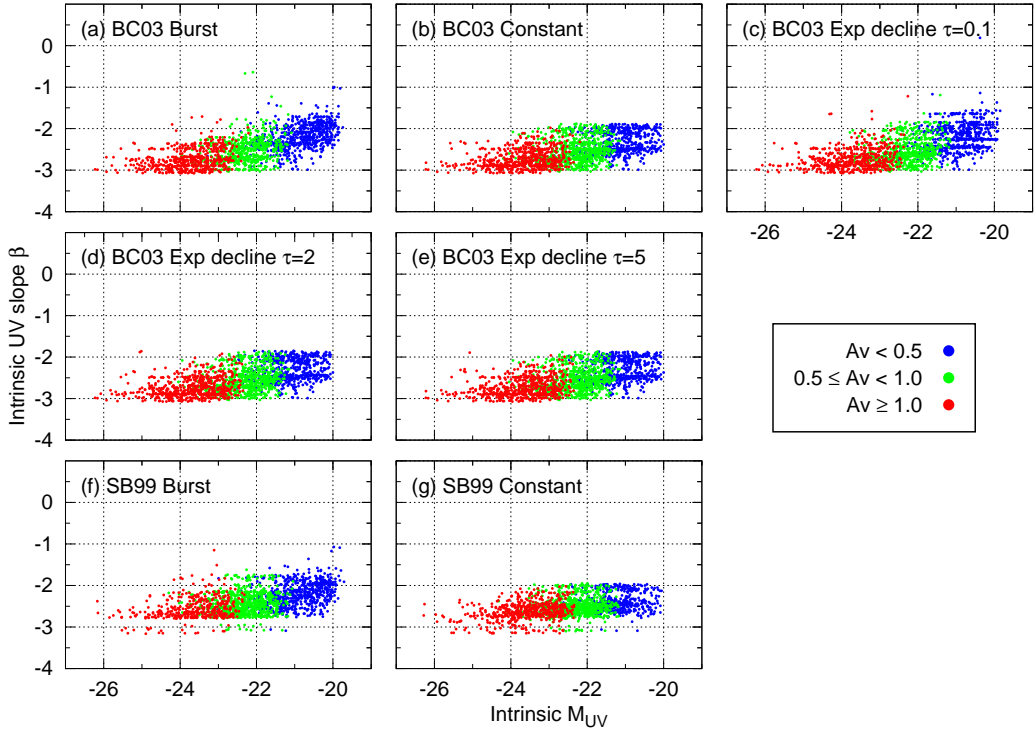}
  \end{center}
  \caption{ Intrinsic UV slope $\beta$ distribution for fixed SFH models. The top three panels and middle two panels show the results of the BC03 model templates and bottom left two panels show the results of the SB99 model templates. The SFHs used for the SED fitting analysis are labeled in each panel: (a) instantaneous burst, (b) continuous constant, (c) exponentially decline with $\tau = 0.1\>$Gyr, (d) exponentially decline with $\tau = 2\>$Gyr, (e) exponentially decline with $\tau = 5\>$Gyr, (f) instantaneous burst, and (g) continuous constant from top to bottom panels. In all of the panels, the best-fit age values for the individual objects are expressed by blue, green, and red color-coding, which indicate the best-fit dust attenuation value of Av $<$ 0.5, 0.5 $\leq$ Av $<$ 1.0, and Av $\geq$ 1.0, respectively. } \label{fig13}
\end{figure*}

\begin{figure*}
  \begin{center}
    \includegraphics[width=130mm]{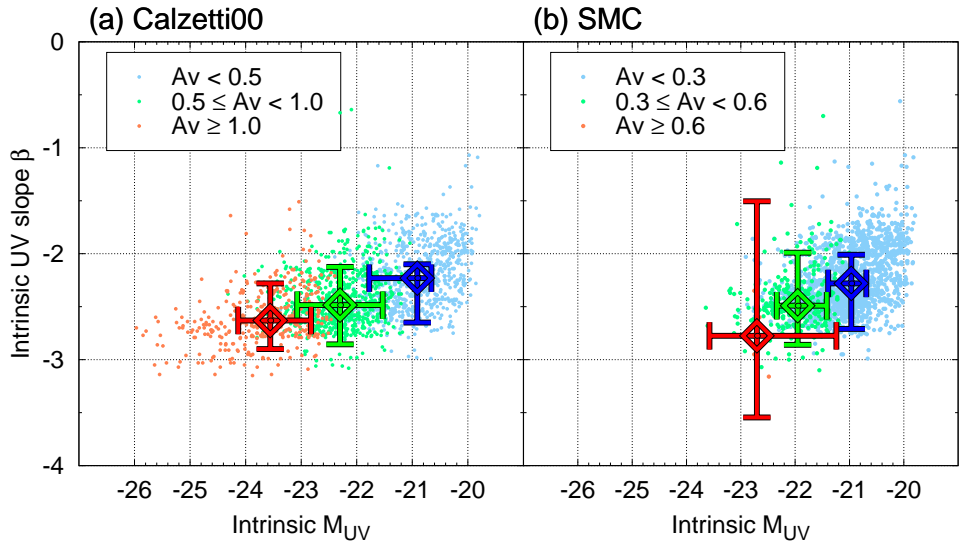}
  \end{center}
  \caption{ Comparison of $\beta_{int}$-$M_{UV,int}$ relation obtained from the \authorcite{Calz00} attenuation law (left) and SMC attenuation law (right). The left panel is same as the left panel of figure \ref{fig11}. The right panel shows the case of the SMC attenuation law for dust extinction curve in the SED fitting analysis, and the blue, green, and red color-coding represent the objects with Av $<$ 0.3, 0.3 $\leq$ Av $<$ 0.6, and 0.6 $\leq$ Av, respectively. The large diamonds with the error bars represent the median value and the median uncertainty. Although the error bars is quite large for the objects with Av $\geq$ 0.6 in the right panel, it is caused from the small number of objects in the sub-sample. } \label{fig14}
\end{figure*}

 First of all, we repeat the SED fitting analysis (1) by fixing SFH parameter and (2) by using SMC attenuation law for dust extinction curve from \citet{Pre84} and \citet{Bouch85}. Figures \ref{fig13} and \ref{fig14} show the result of the case (1) and (2), respectively. The figures show the $\beta_{\mathrm{int}}$--$M_{\mathrm{UV,int}}$ relation, and the fixed SFH parameter or dust extinction curve used in the SED fitting is labeled on the top of each panel. In figure \ref{fig13}, the first and second rows show the results of the BC03 model template and the third row shows the results of the SB99 model template. In all the panels except for the case of the SMC attenuation law, the blue, green, and red points represent the individual objects with the best-fit dust attenuation value of Av $<$ 0.5, 0.5 $\leq$ Av $<$ 1.0, and Av $\geq$ 1.0, respectively. In the case of the SMC attenuation law, the blue, green, and red points represent the objects with Av $<$ 0.3, 0.3 $\leq$ Av $<$ 0.6, and Av $\geq$ 0.6, respectively. In figure \ref{fig14}, the large diamonds with the error bars represent the median value and the median uncertainty for each sub-sample. Although the error bars is quite large for the objects with Av $\geq$ 0.6 in the right panel, it is caused from the small number of objects in the sub-sample.

%--------- Figure13 ---------%
%--------- Figure13 ---------%

 From figure \ref{fig13}, we find that the global trend of the $\beta_{\mathrm{int}}$--$M_{\mathrm{UV,int}}$ relation does not significantly change among SFH parameters, which supports our interpretation described in section \ref{S5s1IBM}. We note that the $\beta_{\mathrm{int}}$ value has discrete values and makes discrete sequences, especially in the panel (c). It is attributed to the age step of the prepared model template in the SED fitting, and the more large number of the age step will dilute the discrete sequences. However, it is not critical when taking account of the moderate uncertainty in photometry.
In brief, the effect of dust attenuation significantly distorts the $\beta_{\mathrm{int}}$--$M_{\mathrm{UV,int}}$ relation, which is probably positive, and then the $\beta$--$M_{\mathrm{UV}}$ relation results in the negative $\beta_{\mathrm{obs}}$--$M_{\mathrm{UV,obs}}$ relation reported by the previous works. In $-22 \lesssim M_{\mathrm{UV,obs}} \lesssim -20$, however, the $\beta_{\mathrm{obs}}$ value seems to be constant to the $M_{\mathrm{UV,obs}}$ value (constant $\beta_{\mathrm{obs}}$--$M_{\mathrm{UV,obs}}$ relation) due to the existence of dusty active star-forming population.

%--------- Figure14 ---------%
%--------- Figure14 ---------%

 Figure \ref{fig14} shows that the best-fit Av value from the SMC attenuation law becomes much smaller than that from the \authorcite{Calz00} attenuation law because the slope of the dust extinction curve of the SMC is much steeper than that of the \authorcite{Calz00}. Consequently, we cannot identify the intrinsically active star-forming galaxies which show the high dust attenuation (Av $>$ 1.0), blue $\beta_{\mathrm{int}}$ value ($\beta_{\mathrm{int}} < -2.42$), and red $\beta_{\mathrm{obs}}$ value ($\beta_{\mathrm{obs}} > -1.7$), although we can again find that the intrinsic $\beta$ value slightly increases with the intrinsic $M_{\mathrm{UV}}$ value. Actually, recent works of Atacama Large Millimeter/submillimeter Array (ALMA) observations report that the SMC dust attenuation law is appropriate for normal star forming galaxies at high redshift (e.g., \cite{Cpk15}; \cite{Bouw16}). On the other hand, as discussed in section \ref{S5s4iasfg}, the \authorcite{Calz00}-like attenuation law is partly required to reproduce the results of the Submillimeter Common User Bolometer Array 2 (SCUBA2) from \citet{Copp15} and \citet{Kopr18}.

%%%%%%%% 5.3 zJK-diagram %%%%%%%%
\subsection{ \textit{zJK}-diagram } \label{S5s3zjkd}

 In this section, we compare the observed color of the \textit{z'JK} band photometry with the predicted color which is estimated from the model simulation. Since our sample tends to have a larger photometric error in the broad-band filters at longer wavelength owing to the depth of the imaging data, the weight of the broad-band filters at longer wavelength becomes smaller than the opposite in the SED fitting analysis. It is possible that the photometry of the \textit{z'JHK} band filters does not have a considerable constraint on the best-fit SED. We therefore focus on the observed color of the \textit{z'JK} band photometry, and more directly compare the observed value with the predicted value in the color--color space.

 For the model simulation, we calculate the color of the two SFH model templates with some condition: One is the BC03 Instantaneous Burst model (hereafter IB), and the other is the BC03 Continuous Constant star formation model (hereafter CSF). We consider that the IB and CSF SFH model is most opposite case in star formation activity, and the models are helpful to interpret the observed results. For the sake of simplicity, we fix the metallicity value with $Z = 0.2 Z_{\solar}$ and the redshift value with \textit{z} $=$ 3.5 and 4.5. In order to clarify the variation of the colors depending on the dust and age, we calculate the colors of the IB and CSF model templates with (a) the fixed age but the variable dust ranging from 0.0 to 3.0, and (b) the variable age ranging from 10$\>\mathrm{Myr}$ to 15.0$\>\mathrm{Gyr}$ but the fixed dust.

%--------- Figure15 ---------%
\begin{figure*}
  \begin{center}
    \includegraphics[width=140mm]{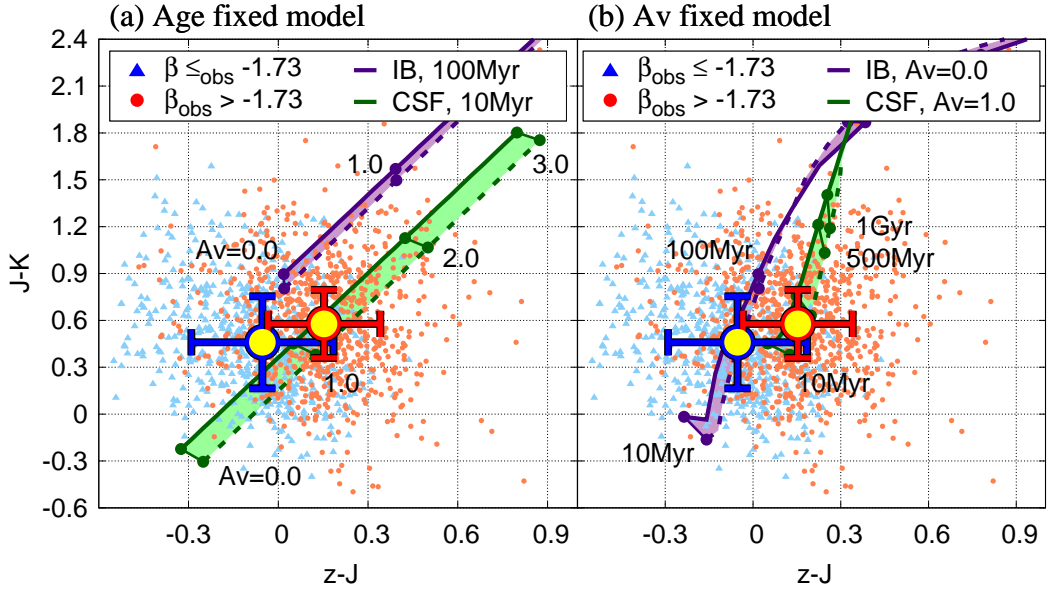}
  \end{center}
  \caption{ $z - J$ vs. $J - K$ color diagram. The blue filled triangles and the red filled circles denote the objects with $\beta_{\mathrm{obs}} \leq -1.73$ and $\beta_{\mathrm{obs}} > -1.73$ in our sample, respectively. The blue and red large circles with the error bars denote the median value and median uncertainty in the observed colors. (a): the green lines represent the CSF model template with age $=$ 10$\>\mathrm{Myr}$, and the purple lines represent the IB model template with age $=$ 100$\>\mathrm{Myr}$. The solid and dashed lines represent the case of \textit{z} $=$ 3.5 and 4.5, respectively, and the space between the lines is filled with the shaded area. The solid circles on each line indicate the dust attenuation value of Av $=$ 0.0, 1.0, 2.0, and 3.0 from bottom left to top right, and the given value is labeled beside the circles. (b): Same as panel (a), but the green lines represent the CSF model template with Av $=$ 1.0, and the purple lines represent the IB model template with Av $=$ 0.0. The solid circles on each line indicate the age value of 10$\>\mathrm{Myr}$, 100$\>\mathrm{Myr}$, 500$\>\mathrm{Myr}$, and 1$\>\mathrm{Gyr}$ from bottom left to top right, and the given value is labeled beside the circles. We note that we omit the label of 100$\>\mathrm{Myr}$ for the green line. } \label{fig15}
\end{figure*}
%--------- Figure15 ---------%

 Figure \ref{fig15} shows the $z - J$ vs. $J - K$ color--color diagram. The vertical axis is the $z - J$ color and the horizontal axis is the $J - K$ color. The $J - K$ color can trace the Balmer break of galaxies at \textit{z} $\sim$ 4 and the $z - J$ color represents the observed UV slope $\beta$. The blue filled triangles and the red filled circles denote the objects with $\beta_{\mathrm{obs}} \leq -1.73$ and $\beta_{\mathrm{obs}} > -1.73$ in our sample, respectively. In this figure, we only use the objects satisfying all the following condition, $z' > 3\,\sigma$, $J > 3\,\sigma$, and $K > 3\,\sigma$, so as to calculate the reliable colors. The blue and red large circles with the error bars denote the median value and median uncertainty in the observed colors. In the left panel (a), the green lines represent the CSF model template with age $=$ 10$\>\mathrm{Myr}$, and the purple lines represent the IB model template with age $=$ 100$\>\mathrm{Myr}$. The solid and dashed lines represent the case of \textit{z} $=$ 3.5 and 4.5, respectively, and the space between the lines is filled with the shaded area. The solid circles on each line indicate the dust attenuation value of Av $=$ 0.0, 1.0, 2.0, and 3.0 from bottom left to top right, and the given value is labeled beside the circles. In the right panel (b), the green lines represent the CSF model template with Av $=$ 1.0, and the purple lines represent the IB model template with Av $=$ 0.0. The solid and dashed lines represent the case of \textit{z} $=$ 3.5 and 4.5, respectively, and the space between the lines is filled with the shaded area. The solid circles on each line indicate the age value of 10$\>\mathrm{Myr}$, 100$\>\mathrm{Myr}$, 500$\>\mathrm{Myr}$, and 1$\>\mathrm{Gyr}$ from bottom left to top right, and the given value is labeled beside the circles. We note that we omit the label of 100$\>\mathrm{Myr}$ for the green line in the panel (b) since the corresponding point is placed under the median value and cannot be seen.

 The figure indicates that the observed distribution of the $\beta_{\mathrm{obs}} > -1.73$ sub-sample tends to be reproduced by the star-forming, dusty, and very young-age, that is bluer $\beta_{\mathrm{int}}$, population. Although we only show the extreme and slightly arbitrary cases in the figure, we can deduce the other possibilities from the examples such as star-forming, less dusty, and middle-age population. However, when we take the other possibilities into consideration, the above interpretation is not changed because the direction of the increase in age and dust is different. We consider that the observed $J - K$ color of the $\beta_{\mathrm{obs}} > -1.73$ sub-sample is not sufficiently red, and thus the middle-age and old-age population is not preferred in the SED fitting analysis. The observed distribution of the $\beta_{\mathrm{obs}} \leq -1.73$ sub-sample tends to be reproduced by the less star-forming, less dusty, and young-age population, although the sub-sample can be also reproduced by the star-forming, less dusty, and middle-age population.  We note that there are some outliers in our sample, but most of them have a lower signal to noise ratio ($S/N \sim 3$--$5$) in \textit{J} and/or \textit{K} band than the other objects. In summary, the interpretation from the \textit{z'JK} color--color diagram is consistent with the interpretation from our SED fitting analysis, and therefore a part of star-forming galaxies at \textit{z} $\sim$ 4 in our sample is indeed classified as dusty star-forming population.

%%%%%%%% 5.4 intrinsically active star forming galaxies %%%%%%%%
\subsection{ Expected features of most active star-forming galaxies at \textit{z} $\sim$ 4 } \label{S5s4iasfg}

 Last of this paper, we show two estimation for the IR features of the active star-forming galaxies at \textit{z} $\sim$ 4: One is the luminosity ratio of IR to UV so called IRX, and the other is the flux density at observed-frame 850$\>\micron$, $S_{850}$. Our sample does not have the rest-frame IR information for the individual objects and therefore we use the approximate conversion. For estimating the IRX value, we apply the empirical conversion between ${\rm IRX_{TIR-FUV}}$ and ${\rm A_{FUV}}$ for low-\textit{z} galaxies reported by \citet{Burg05}: ${\rm A_{FUV}} = -0.028[{\rm log_{10}}L_{\rm TIR}/L_{\rm FUV}]^3 + 0.392[{\rm log_{10}}L_{\rm TIR}/L_{\rm FUV}]^2 + 1.094[{\rm log_{10}}L_{\rm TIR}/L_{\rm FUV}] + 0.546$. For estimating the $S_{\mathrm{850}}$ value, we first calculate the total (bolometric) IR luminosity by utilizing the not dust-corrected FUV luminosity and the IRX value, and then we convert the total IR Luminosity into the flux density at observed-frame 850$\>\micron$. In the conversion, we use the modified blackbody $+$ power-law formula as the dust thermal emission and the total IR luminosity is estimated by integrating the modeled spectrum from 8$\>\micron$ to 1000$\>\micron$ in the rest-frame. The formula is,
\begin{equation}
S (\nu, T_{d}) \propto \left\{ 
\begin{array}{ll}
\frac{\nu^{\beta}\nu^{3}}{e^{h\nu/kT_{d}}-1} & (\nu \leq \nu_{c}); \\
\nu^{-\alpha} & (\nu > \nu_{c}),
\end{array}
\label{eq2}
\right.
\end{equation}
where $S(\nu, T_{d})$ is the flux density at $\nu$ for a dust temperature $T_{d}$ in the units of Jy and $\beta$ is a dust emissivity index. The connecting frequency, $\nu_{c}$, is calculated from,
\begin{equation}
\Biggl. \frac{dS}{d\nu} \Biggr|_{\nu=\nu_{c}} = -\alpha.
\label{eq3}
\end{equation}
For the sake of simplicity, we fix all the above parameters and the source redshift referring to \citet{Copp15}: the dust temperature of $T_{d} = 38\>$K, the dust emissivity index of $\beta_{\mathrm{dust}} = 1.5$, the power-law index of $\alpha = 1.7$, and the source redshift of $z = 3.87$. We emphasize that the cautious treatment is required for the comparison between our result and the previous results presented in this paper.

%--------- Figure16,17,18 ---------%
\begin{figure*}
  \begin{center}
    \includegraphics[width=52mm]{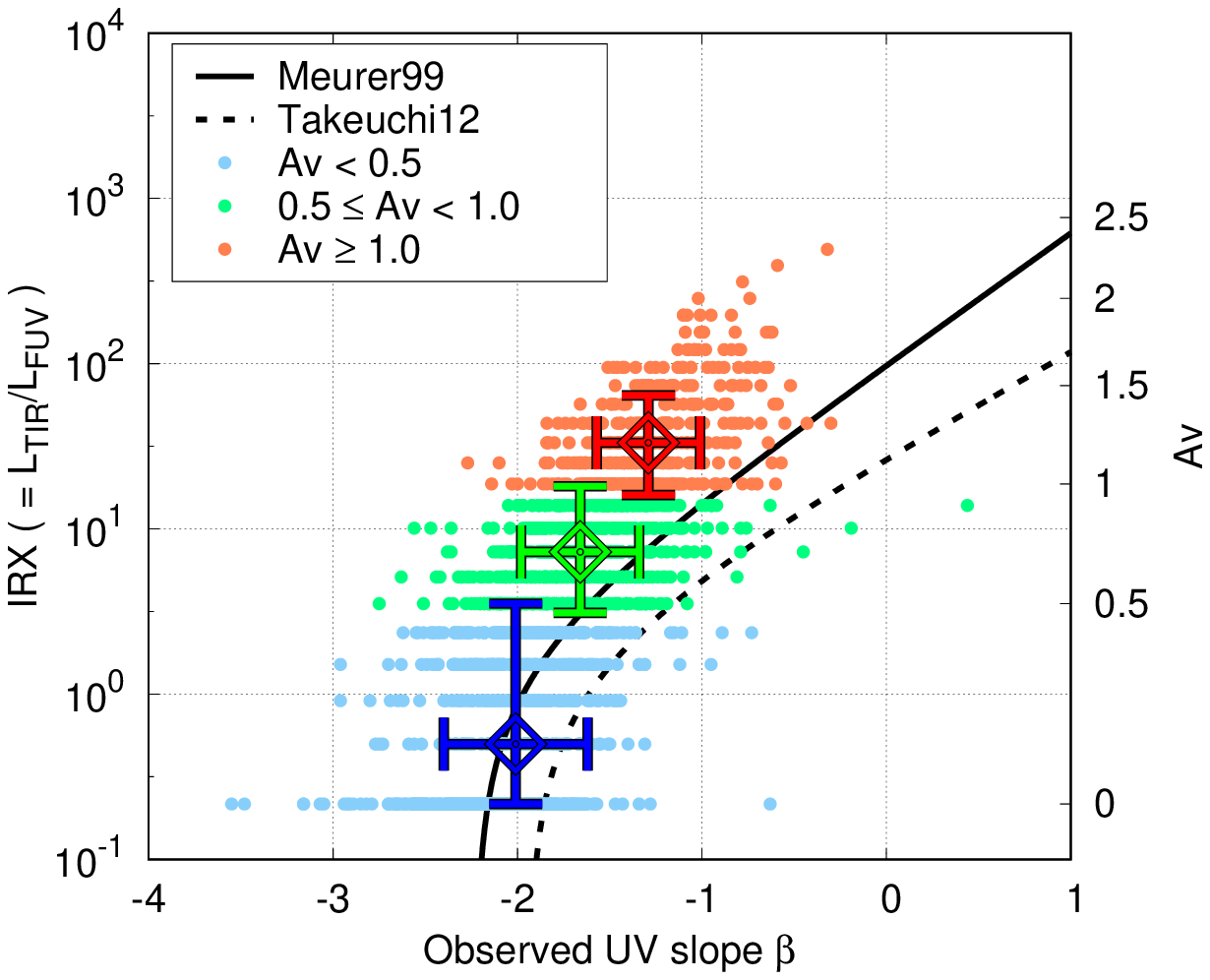}
    \includegraphics[width=52mm]{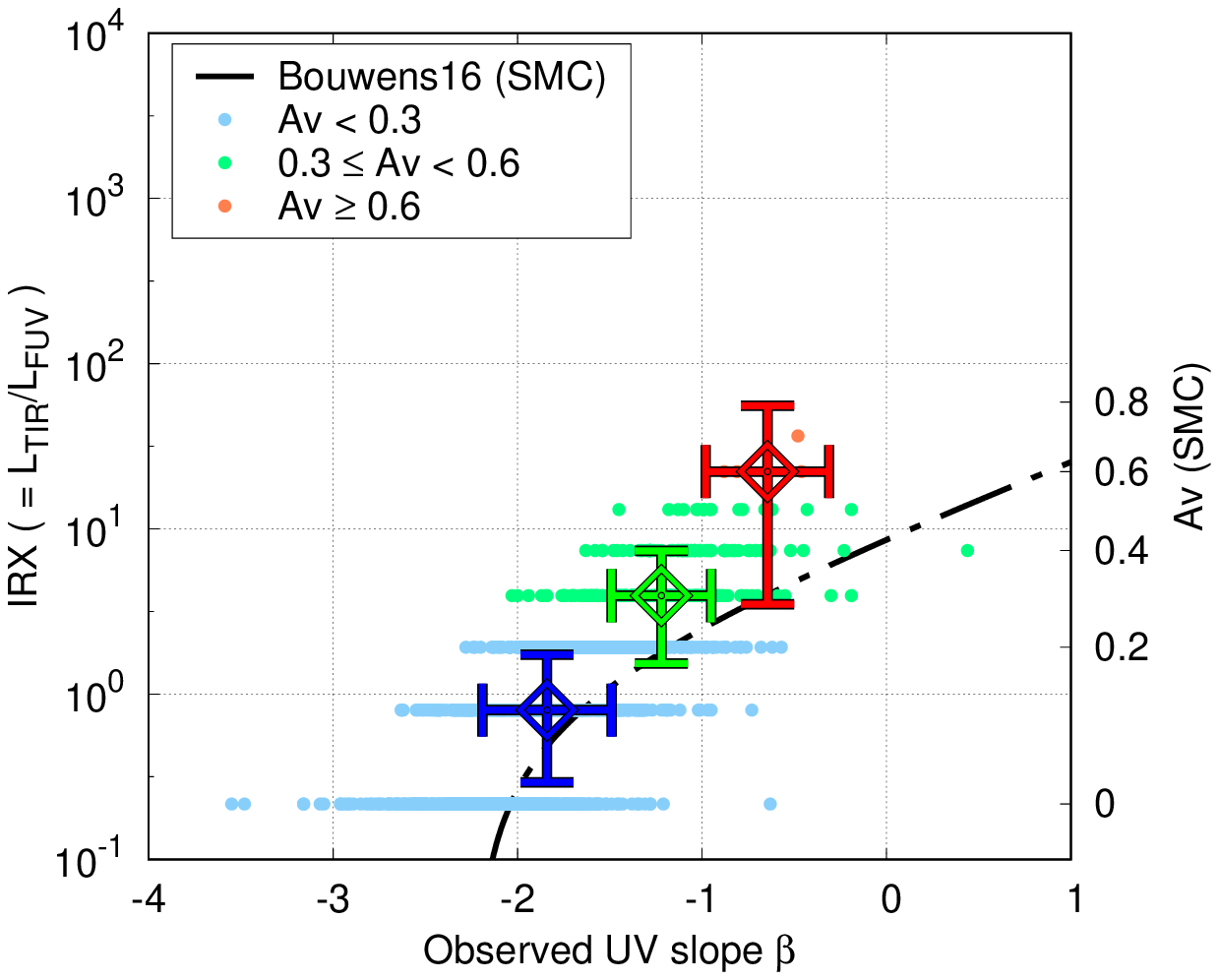}
    \includegraphics[width=52mm]{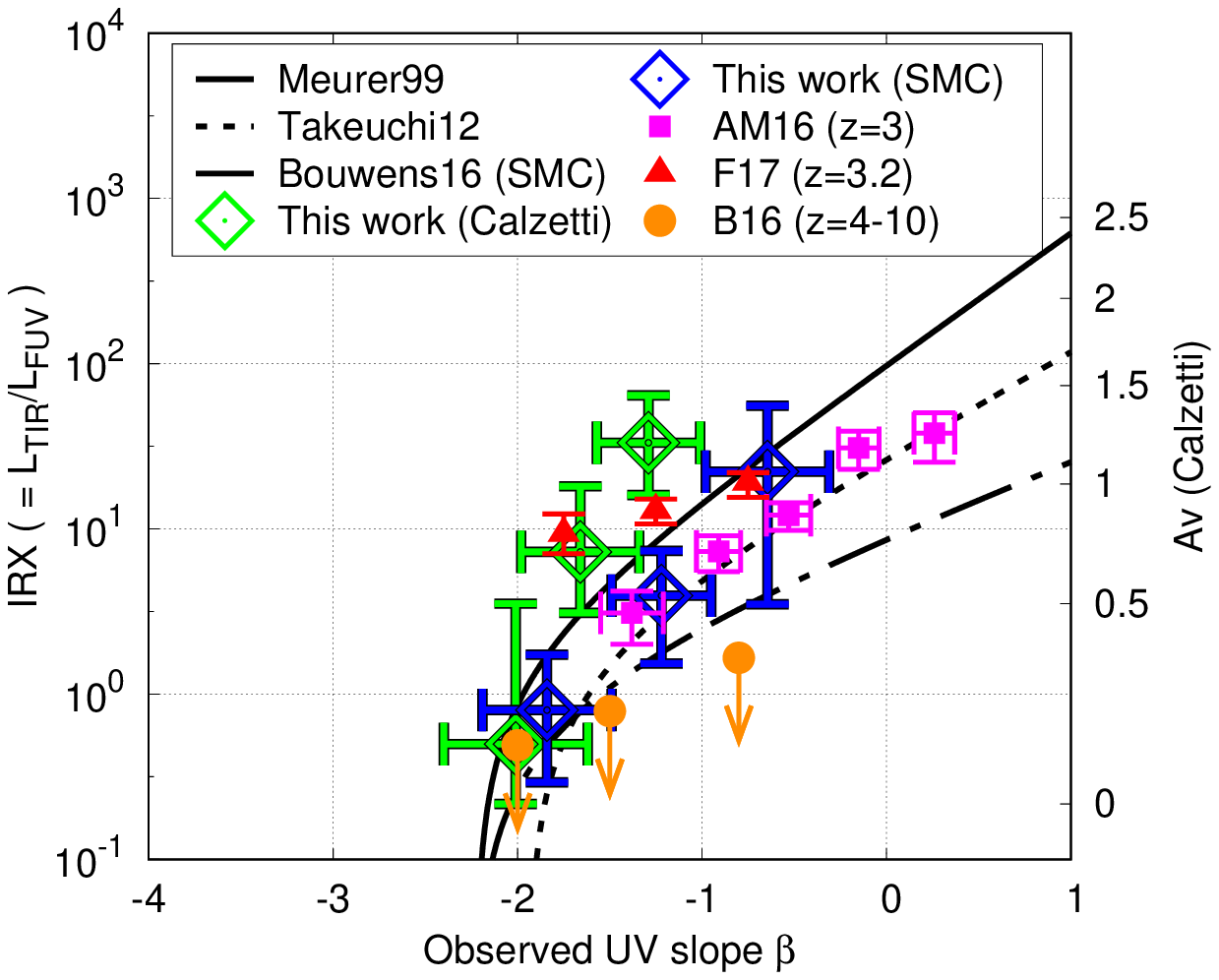}
  \end{center}
  \caption{ Predicted distribution of IRX--$\beta$ relation in the case of the \authorcite{Calz00} attenuation law (left) and SMC attenuation law (middle), and the comparison with previous works (right). The IRX value is estimated by the empirical conversion between $\mathrm{IRX_{TIR-FUV}}$ and $\mathrm{A_{FUV}}$ reported by \citet{Burg05}. The blue, green, and red points in the left panel (middle panel) represent the best-fit dust value of Av $\leq$ 0.5 (0.3), 0.5 (0.3) $<$ Av $\leq$ 1.0 (0.6), and Av $\geq$ 1.0 (0.6), respectively. The large square with the error bars denote the median value and median uncertainty. In the right panel, we show the median value of our result, and the magenta squares, red triangles, and orange circles represent the results from \citet{Alva16}, \citet{Fuda17}, and \citet{Bouw16}, respectively. In the left and right panels, the black solid and dashed lines show the relation based on the \authorcite{Calz00} attenuation law from \citet{Meur99} and \citet{Take12}, respectively. In the middle and right panels, the black dot-dashed line shows the relation based on the SMC attenuation law from \citet{Bouw16}.} \label{fig161718}
\end{figure*}
%--------- Figure16,17,18 ---------%

 Figure \ref{fig161718} shows the IRX--$\beta$ relation obtained from the \authorcite{Calz00} attenuation law (left) and SMC attenuation law (middle). The vertical axis is the predicted IRX value, and the horizontal axis is the observed UV slope $\beta$. The small dots represent each object, and the color-coding is same as the figure \ref{fig14}. The large blue, green, and red square with the error bars represent the median value and median uncertainty for each sub-sample. In the right panel, we show the median values of our result and the previous works from \authorcite{Alva16}(2016: AM16, magenta square), \authorcite{Fuda17}(2017: F17, red triangle), and \authorcite{Bouw16}(2016: B16, orange circle). The sample of AM16 is LBGs at \textit{z} $\sim$ 3 in the COSMOS field and the IR luminosity is obtained from the stacked image of the Herschel and AzTEC. The sample of F17 is massive star-forming galaxies at \textit{z} $\sim$ 3.2 in the COSMOS field, which are distributed on the main-sequence of star formation, and the IR luminosity is obtained from the stacked image of the ALMA. We note that both the samples consist of the relatively more massive ($\mathrm{M_{*} \gtrsim 10^{10} M_{\solar}}$) and lower redshift LBGs compared with our sample. The sample of B16 is LBGs at \textit{z} $=$ 4--10 in the Hubble Ultra Deep Field, and the IR Luminosity is obtained from the stacked image of the ALMA. For B16, the data points in this panel represent the 2$\,\sigma$ upper limit of the formal uncertainty for the $\mathrm{M_{*} < 10^{9.75} M_{\solar}}$ sample described in table 13 of their paper, and thus their sample is relatively less massive (and possibly higher redshift) LBGs compared with our sample. The black solid and dashed lines show the relation based on the \authorcite{Calz00} attenuation law from \citet{Meur99} and \citet{Take12}, respectively. The black dot-dashed line shows the relation based on the SMC attenuation law from \citet{Bouw16}.

 In the case of the \authorcite{Calz00} attenuation law (left panel), our sample shows the systematically bluer UV slope $\beta$ and the systematic offset becomes larger at the larger IRX value. According to previous works for lower redshift star-forming galaxies (e.g., \cite{Redd06}; \cite{Hein13}; \cite{Oteo13}; \cite{Alva16}), normal star-forming galaxies are distributed along the IRX--$\beta$ relation, and IR luminous galaxies such as Luminous InfraRed Galaxies (LIRGs; $L_{\mathrm{TIR}} > 10^{11}\,L_{\solar}$) or Ultra Luminous InfraRed Galaxies (ULIRGs; $L_{\mathrm{TIR}} > 10^{12}\,L_{\solar}$) are distributed above the IRX--$\beta$ relation. The offset of our red points implies the presence of IR excess galaxies at at \textit{z} $\sim$ 4 such as local LIRG/ULIRGs although the systematic shift can be attributed to the uncertainty of IRX which comes from the conversion from $A_{\mathrm{FUV}}$ to $\mathrm{IRX_{TIR−-FUV}}$ and/or the failure of the SED fitting analysis. In the case of the SMC attenuation law (middle panel), our sample also shows the systematically bluer UV slope $\beta$ especially at the larger IRX value. Most of our sample, however, show the moderate IRX value (IRX $\leq$ 10), and we find a few IR excess galaxies in our sample. In conclusion, our sample indicates the presence of the IR excess galaxies at \textit{z} $\sim$ 4.

 When comparing the previous works (right panel in figure \ref{fig161718}), our sample from \authorcite{Calz00} attenuation law tends to have the bluer UV slope $\beta$ at the larger IRX value than all the previous works while our sample from SMC attenuation law is comparable to the those of AM16 and F17. Our results from both attenuation law are not consistent with the result of B16. We note that the difference in the stellar mass of the sample is critical for the IRX--$\beta$ relation since both the IRX and $\beta$ values depend on the stellar mass (e.g., \cite{Alva16}; \cite{Bouw16}; \cite{Fink12}; \cite{Fuda17}), and we consider that the inconsistency between our results and B16 is attributed to the difference in the stellar mass. The red data point from F17 at $\beta \sim -1.7$ and IRX $\sim$ 10 (most left side point) is comparable to our result from \authorcite{Calz00} attenuation law although the other data point from F17 is comparable to that from SMC. The authors mention that the most left side point is uncertain because of the small sample size of the bin. Therefore, our result from SMC attenuation law is not inconsistent with the previous works.

%--------- Figure19,20 ---------%
\begin{figure*}
  \begin{center}
    \includegraphics[width=60mm]{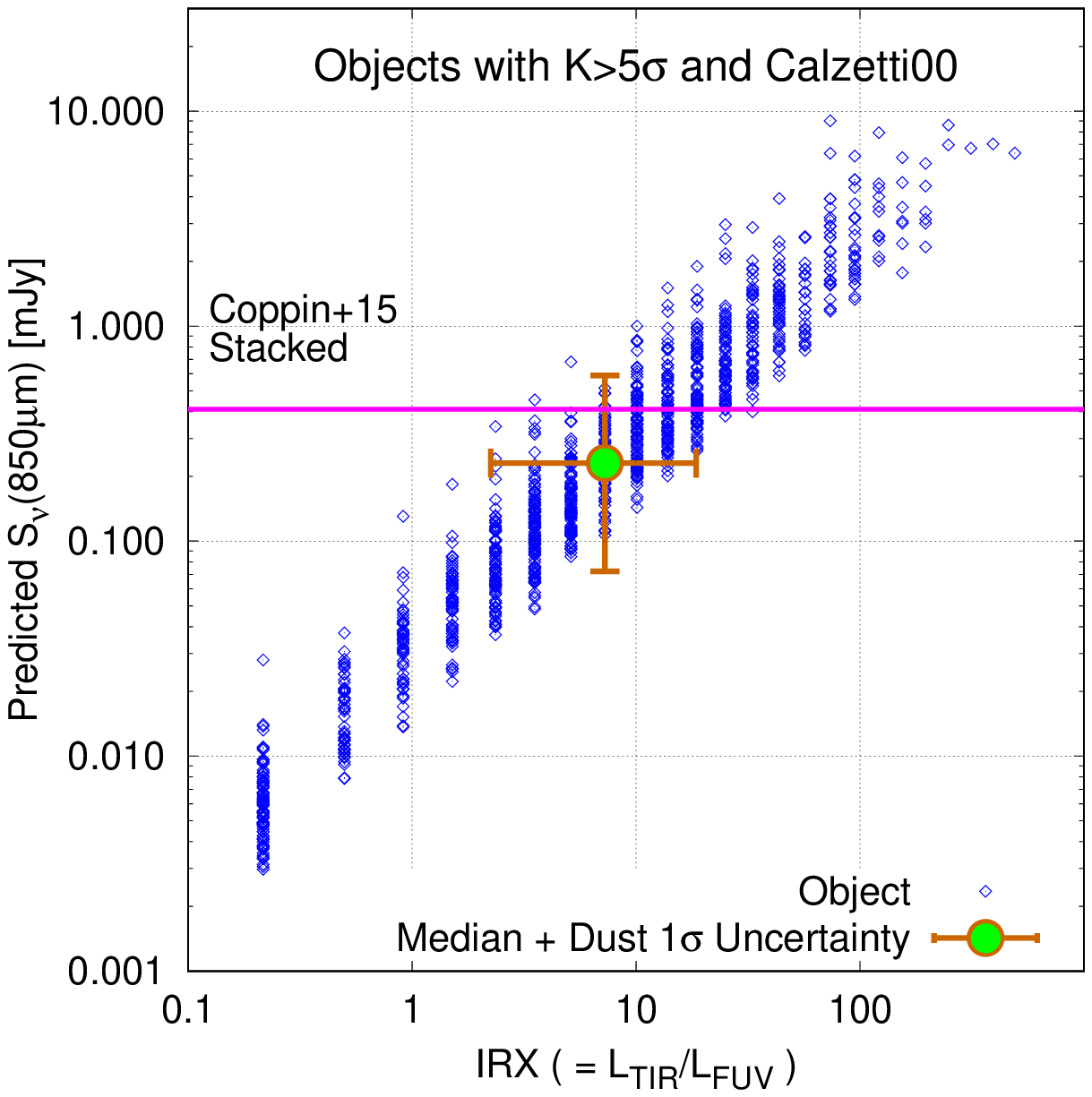}
    \includegraphics[width=60mm]{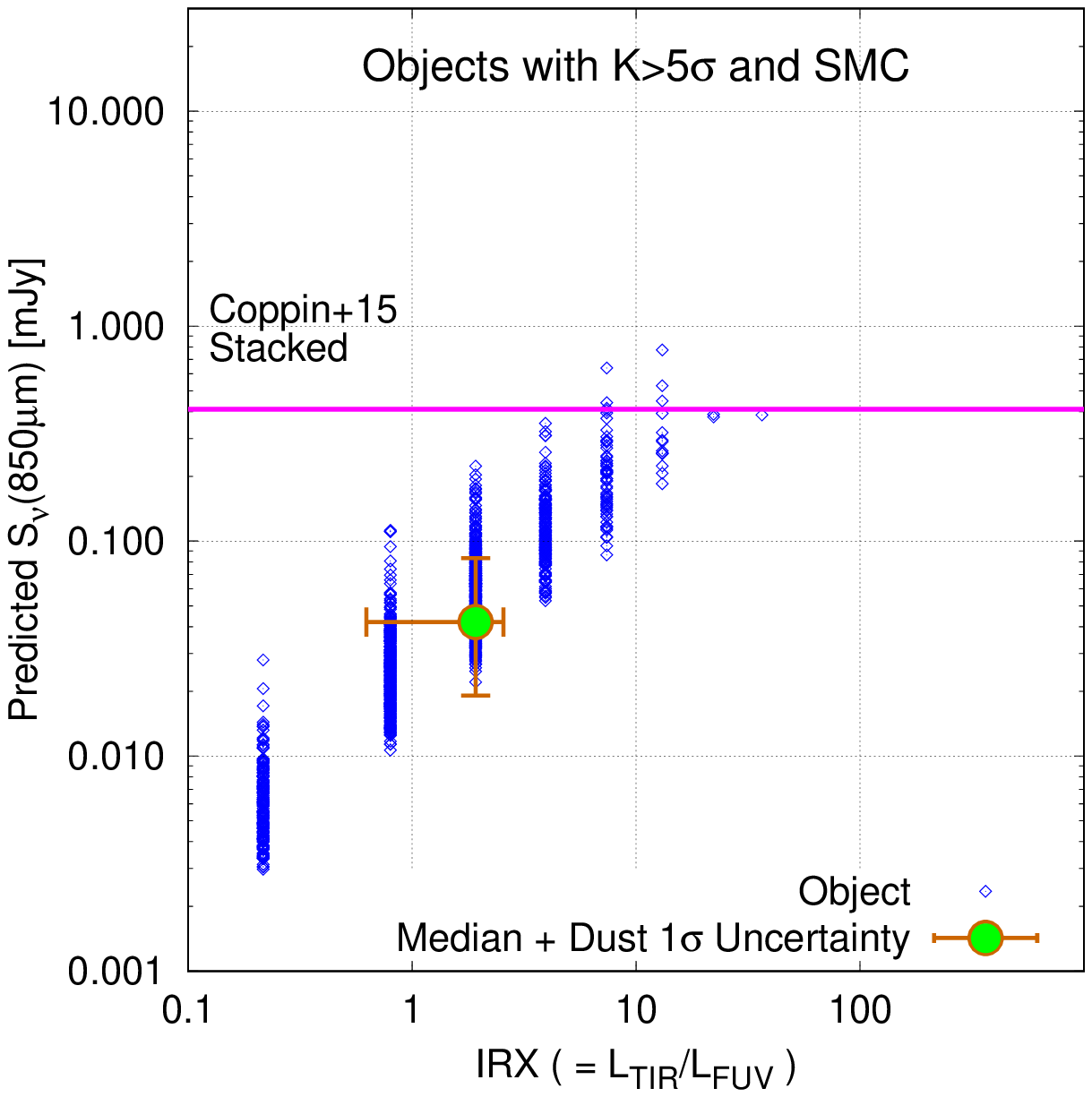}
  \end{center}
  \caption{ Predicted flux density at observed-frame 850$\>\micron$ for our sample. The left panel shows the case of the \authorcite{Calz00} attenuation law, and the right panel shows the case of the SMC attenuation law. The vertical axis is the predicted $S_{\mathrm{850}}$ value and the horizontal axis is the predicted IRX value. The blue open diamonds represent the individual objects detected at $> 5\,\sigma$ level in \textit{K} band photometry. The green filled circle with the orange error bars denotes the median value and median uncertainty estimated from the uncertainty in Av. The horizontal magenta solid line denotes the flux density measured in \citet{Copp15}. } \label{fig1920}
\end{figure*}
%--------- Figure19,20 ---------%

For the further verification of our result, figure \ref{fig1920} shows the prediction of $S_{\mathrm{850}}$ for the case of the \authorcite{Calz00} attenuation law (Left) and the SMC attenuation law (Right). The vertical axis is the predicted $S_{850}$ value and the horizontal axis is the predicted IRX value. The blue open diamonds represents the individual objects detected at $> 5\,\sigma$ level in \textit{K} band photometry in our sample. The green filled circle with the orange error bars denotes the median value and median uncertainty of the whole sample. The uncertainty is also estimated from the uncertainty in Av. The horizontal magenta solid line denotes the flux density for the stacked LBG at \textit{z} $\sim$ 4 measured in \citet{Copp15} whose sample is quite similar to our sample. Since the sample in \citet{Copp15} consists of the \textit{K} band detected objects, we only show the $K > 5\,\sigma$ objects in the figure. According to \citet{Copp15}, the flux density measured for the stacked image is $S_{\mathrm{850}} = 0.411 \pm 0.064\>$mJy. We note that the result of \citet{Copp15} is derived from the SED template library constructed by \citet{Swin14} and the modified blackbody $+$ power-law formula is used just for checking the validity of their SED fitting analysis.

 The figure shows that the predicted $S_{\mathrm{850}}$ flux from the SMC attenuation law is insufficient to reproduce the stacking result of \citet{Copp15}, but the result from the \authorcite{Calz00} attenuation law is consistent with the stacking result. This comparison indicates that a part of \textit{z} $\sim$ 4 LBGs are indeed significantly dust attenuated and there must be IR luminous star-forming galaxies in our sample. Alternatively, at least the SMC attenuation law is unsuitable for high-\textit{z} and \textit{K}-detected LBGs. However, the difference between \citet{Copp15} result and ours can be due to the fact that the stacking result is the average weighted by luminosities while our median values are not. Since the red LBGs in our sample can be easily detected and measured by using the ALMA, future ALMA observations for individual detection will potentially solve the discrepancy.

 We consider the possible interpretation of our optical/NIR-based IRX--$\beta$ relation. The IRX--$\beta$ relation is expressed as $\log_{10} \mathrm{IRX} = \log_{10}(10^{0.4*\mathrm{c1}*(\beta - \beta_{0})} - 1.0) + \mathrm{c2}$ where c1, c2, and $\beta_{0}$ are a constant value. The c1 value is the slope of the relation between dust attenuation Av and the observed UV slope $\beta$, $d \mathrm{Av}/d \beta$, which is specified by the dust extinction curve. The c2 value represents the bolometric correction because the observed UV and IR Luminosity is not the representative value and we need the correction factor for the observed values. The $\beta_{0}$ value is the intrinsic UV slope $\beta$ as investigated in this paper. In short, the IRX--$\beta$ relation assumes that the extinction curve and the stellar population hidden by dust does not vary significantly with the physical quantities of the star-forming galaxies.
 In the previous works for the IR-based IRX--$\beta$ relation, by using the fixed $\beta_{0}$ value ($\sim -2.2$), the authors discuss the suitable extinction curve for reproducing the IRX--$\beta$ relation seen in high redshift galaxies (e.g., \cite{Cpk15}; \cite{Alva16}; \cite{Bouw16}). \citet{Redd18} explain the the IRX--$\beta$ relation of the \textit{z} $\sim$ 2 galaxies by using the SMC attenuation law and the more bluer $\beta_{0}$ value ($\sim -2.6$), which is derived from the recent stellar population synthesis model. Moreover, \citet{LeeKS12} and \citet{Redd12} discuss the variation of the extinction curve according to the observed UV magnitude and the age of star-forming galaxies.
 From our analysis, assuming a certain dust extinction curve, the observed properties are not represented by the IRX--$\beta$ relation with the fixed $\beta_{0}$ value, and it is required that there is the variation of the intrinsic $\beta$ value or the variation of the extinction curve, or both, depending on the physical quantities of the star-forming galaxies. The prediction of the $S_{850}$ flux indicates that our sample is expected to include the highly dust attenuated and IR luminous galaxies which are explained by the \authorcite{Calz00} attenuation law. Therefore, while the less dusty galaxies can be characterized by either attenuation law of \authorcite{Calz00} and SMC, the highly dust attenuated galaxies are most likely characterized by the \authorcite{Calz00} attenuation law and the bluer intrinsic $\beta$ value. Although it is difficult to confirm the variation from our results, there seems to be the variation of the intrinsic $\beta$ value or the extinction curve according to the physical quantities of the star-forming galaxies.

%%%%%%%% 6. Conclusion %%%%%%%%
\section{Conclusion} \label{S6Conc}

 In this work, we investigate the UV slope $\beta$ and stellar population of bright star-forming galaxies at \textit{z} $\sim$ 4 in the SXDS field which is the wide-area and deep survey field. We use the imaging data of Subaru/\textit{BVRi'z'}updated-\textit{z'}, UKIRT/\textit{JHK}, HST/F125WF160W, and Spitzer/3.6$\>\micron$ 4.5$\>\micron$, and we construct the sample of star-forming galaxies at \textit{z} $\sim$ 4 by both Lyman Break technique and photometric redshift selection. The UV slope $\beta$ is calculated by the simple power-law fit, and the stellar population is estimated from the optical and NIR photometry thorough the SED fitting analysis. Consequently, we find a sign that some star-forming galaxies, which experience on-going active star formation and suffer heavy dust attenuation, really exist in the \textit{z} $\sim$ 4 universe. We list our main results below.

\begin{itemize}
\item
  There seems to be little dependence of the observed UV slope $\beta$ on the observed UV absolute magnitude $M_{\mathrm{UV}}$ in the range of $-22.0 \lesssim M_{\mathrm{UV}} \lesssim -20.0$ although the dynamic range of $M_{\mathrm{UV}}$ is limited. The slope of the $\beta$--$M_{\mathrm{UV}}$ relation is $-$0.02 $\pm$ 0.02, and it is more shallower than the previous studies for similar redshift but fainter LBGs ($-$0.13 $\pm$ 0.02 from \cite{Bouw14} and $-$0.10 $\pm$ 0.03 from \cite{Kurc14}). 

\item
  For investigating the dependence of the UV slope $\beta$ on the dust attenuation, age, metallicity, and SFH, we calculate the \textit{intrinsic} (dust-corrected) UV slope, $\beta_{\mathrm{int}}$, and \textit{intrinsic} UV absolute magnitude, $M_{\mathrm{UV,int}}$, by using the results of the SED fitting analysis. The star-forming galaxies with the bluest $\beta_{\mathrm{int}}$ and brightest $M_{\mathrm{UV,int}}$ value are the dusty star-forming population which is observed with the red $\beta_{\mathrm{obs}}$ value. The dusty star-forming population has $\beta_{\mathrm{obs}} > -1.73$, $\mathrm{Av} \geq 1.0$, $\beta_{\mathrm{int}} \leq -2.42$, and SFR $\gtrsim$ a few $\times\ 10^{2}\,\mathrm{M_{\solar}}\>\mathrm{yr^{-1}}$, and we see the flat $\beta_{\mathrm{obs}}$--$M_{\mathrm{UV,obs}}$ distribution due to such population.

\item
  We find the intersection point of the $\beta_{\mathrm{int}}$--$M_{\mathrm{UV,int}}$ relation and the $\beta_{\mathrm{obs}}$--$M_{\mathrm{UV,obs}}$ relation by extrapolating our relation toward the fainter magnitude range. The intersection point represents the position of the appearance of nearly dust-free population, and it is at $\beta = -1.94$ and $M_{UV} = -18.88$ which is close to the break point of $\beta_{\mathrm{obs}}$--$M_{\mathrm{UV,obs}}$ relation reported by \citet{Bouw14}.

\item
  Our result does not depend on the SFHs used in the SED fitting analysis. However, our result depends on the assumption of the attenuation law. The best-fit dust attenuation value assuming the SMC attenuation law is found to be smaller than that obtained with the \authorcite{Calz00} attenuation law. The trend that the intrinsic $\beta$ value increases with the intrinsic $M_{\mathrm{UV,int}}$ value appears for both the cases.

\item
  We compare the observed color of the \textit{zJK} broad-band filters with the expected colors. Since the \textit{z-J} color traces the UV slope $\beta$ and the \textit{J-K} color traces the Balmer break of \textit{z} $\sim$ 4 LBGs, we can also infer the stellar population by the observed quantities. The observed color of the $\beta_{\mathrm{obs}} > -1.73$ sub-sample of the \textit{z} $~$ 4 star-forming galaxies is well reproduced by star-forming, dusty, and young-age (blue $\beta_{\mathrm{int}}$) population.

\item
  We estimate the IRX ($= L_{\mathrm{TIR}}/L_{\mathrm{FUV}}$) value and the flux density at observed-frame 850$\>\micron$, $S_{\mathrm{850}}$, from only the optical and NIR imaging data. The optical/NIR-based IRX--$\beta$ relation indicates the variation of the intrinsic $\beta$ value or the variation of the dust attenuation law, or both, according to the physical quantities of the star-forming galaxies. The $S_{\mathrm{850}}$ value estimated from the SMC attenuation law is not consistent with the stacking results of \citet{Copp15}, and thus the \authorcite{Calz00} attenuation law is preferable to the \textit{z} $\sim$ 4 intrinsically luminous LBGs.

\item
  Our analysis indicates that a significant fraction of \textit{z} $\sim$ 4 LBGs are the highly dust attenuated and IR luminous population such as ULIRGs/LIRGs. This population has not been recognized very well in the previous analysis but is important in understanding early phase of galaxy formation possibly linking the typical blue LBGs and the further very red sub-mm selected galaxies.

\end{itemize}

%%%%%%%% Acknowledgments %%%%%%%%
\begin{ack}
  We appreciate M. Kajisawa, A. Inoue, K. mawatari, and T. Hashimoto for helpful comments and discussions. 
  This work is mainly based on data collected at Subaru Telescope, which is operated by the National Astronomical Observatory of Japan.
  The UKIDSS project is defined in \citet{Law07}. UKIDSS uses the UKIRT Wide Field Camera (WFCAM; \cite{Cas07}). The photometric system is described in \citet{Hew06}, and the calibration is described in \citet{Hodg09}. The pipeline processing and science archive are described in Irwin et al (2009, in prep) and \citet{Hamb08}. We used UKIDSS data release 10.
  This work is based on observations taken by the CANDELS Multi-Cycle Treasury Program with the NASA/ESA HST, which is operated by the Association of Universities for Research in Astronomy, Inc., under NASA contract NAS5-26555.
  This work is based in part on observations made with the Spitzer Space Telescope, which is operated by the Jet Propulsion Laboratory, California Institute of Technology under a contract with NASA.
  Data analysis was in part carried out on the open use data analysis computer system at the Astronomy Data Center, ADC, of the National Astronomical Observatory of Japan. We used the interactive analysis servers (anam[01-16]), the batch processing servers (bapm[01-06]), the terminal workstations (new-r[01-13]), and the disk space (home and mfst).
  This work was supported by JSPS KAKENHI Grant Number JP26400217.
\end{ack}

%%%%%%%% References %%%%%%%%
%%%
% See the manual for the detail.
%%%

\end{document}